\documentclass[12pt]{article}

\usepackage{scicite}

\usepackage{times}
\usepackage{graphicx}
\usepackage{amssymb}
\usepackage{verbatim, longtable}	% Easily comment out text; Long tables
\usepackage{color, soul}	% Highlighting, use \hl{...}
\usepackage[colorlinks,pdftex,urlcolor=blue,citecolor=blue,linkcolor=blue]{hyperref}	% Fancy links (PDF)
\usepackage{lscape}	% for rotating the longtable
\usepackage{todonotes}	% side notes at the margins, use \todo

\def\aap{{\it Astron.~Astrophys.}~}

\def\apj{{\it Astrophys.~J.}~}

\def\apjs{{\it Astrophys.~J.~Suppl.~Ser.}~}
\def\araa{{\it Annu.~Rev.~Astron.~Astrophys.}~}
\def\mnras{{\it Mon.~Not.~R.~Astron.~Soc.}~}
\def\nat{{\it Nature}~}
\def\sci{{\it Science}~}

\newenvironment{sciabstract}{%
\begin{quote} \bf}
{\end{quote}}

\newcounter{lastnote}

\title{A Universal Scaling for the Energetics of Relativistic Jets From Black Hole Systems} 

\date{}

\begin{document}

\maketitle 

\noindent
R.~S.~Nemmen$^{1*}$, 
M.~Georganopoulos$^{2,1}$, 
S.~Guiriec$^{1}$,
E.~T.~Meyer$^{3,5}$, 
N.~Gehrels$^{1}$, 
R.~M.~Sambruna$^{4}$
\medskip
\begin{enumerate}
\item[1.] NASA Goddard Space Flight Center, Greenbelt, MD 20771, USA
\item[2.] Department of Physics, University of Maryland Baltimore County, 1000 Hilltop Circle, Baltimore, MD 21250, USA
\item[3.] Department of Physics and Astronomy, Rice University, Houston, TX 77005, USA
\item[4.] Department of Physics and Astronomy, George Mason University, MS 3F3, 4400 University Drive, Fairfax, VA 22030, USA
\item[5.] Space Telescope Science Institute, 3700 San Martin Drive, Baltimore, MD 21218, USA
\end{enumerate}
$^*$To whom correspondence should be addressed. E-mail: \href{mailto:rodrigo.nemmen@nasa.gov}{rodrigo.nemmen@nasa.gov}
\\
\todo[inline,color=green!40]{Published in \href{http://www.sciencemag.org/content/338/6113/1445.abstract}{\emph{Science}, 338, 1445 (2012), DOI: 10.1126/science.1227416}. \\
This is the author's version of the work. It is posted here by permission of the AAAS for personal use, not for redistribution.}

\clearpage

\begin{sciabstract}
Black holes generate collimated, relativistic jets which have been observed in gamma-ray bursts (GRBs), microquasars, and at the center of some galaxies (active galactic nuclei; AGN).
How jet physics scales from stellar black holes in GRBs to the supermassive ones in AGNs is still unknown. 
Here we show that jets produced by AGNs and GRBs exhibit the same correlation between the kinetic power carried by accelerated particles and the $\gamma$-ray luminosity, with AGNs and GRBs lying at the low and high-luminosity ends, respectively, of the correlation. 
This result implies that the efficiency of energy dissipation in jets produced in black hole systems is similar over 10 orders of magnitude in jet power, establishing a physical analogy between AGN and GRBs.  
\end{sciabstract}

Relativistic jets are ubiquitous in the cosmos and have been observed in a diverse range of black hole systems spanning from stellar mass ($\sim 10 M_\odot$) to supermassive scales ($\sim 10^5-10^{10} M_\odot$), in particular in the bright flashes of gamma-rays known as GRBs \cite{gehrels09,piran04}, the miniature versions of quasars lurking in our galaxy known as ``microquasars'' \cite{mirabel99} and AGNs \cite{rees84,krolik}. 
Despite decades of observations at almost all wavelengths and considerable theoretical efforts, there are still many aspects of black hole jets which remain mysterious: the mechanism(s) responsible for their formation and the nature of their energetics as well as their high-energy radiation \cite{meier01, nar05}. 
Jets and outflows from supermassive black holes have important feedback effects on scales ranging from their host galaxies to groups and clusters of galaxies \cite{mcnamara07}. Hence, a better understanding of the physics of jets is required in order to have a more complete picture of the formation and evolution of large-scale structures in the universe and the coevolution of black holes and galaxies \cite{sijacki07}.

One outstanding question is how the jet physics scale with mass from stellar to supermassive black holes. Interestingly, there is evidence that jets behave in similar ways in microquasars and radio-loud AGN \cite{marscher02,merloni03,falcke04}. However, a clear connection between AGN and GRBs has not been established yet, although recent work provides encouraging results \cite{wang11,wu11}.

As a first step in understanding how the properties of jets vary across the mass scale, we focus on the energetics of jets produced in AGNs and GRBs. 
Therefore, we searched the literature for published and archival observations that allow us to estimate the jet radiative output and the kinetic power for a sample of black hole systems in which the jet is closely aligned with our line of sight and characterized by a broad range of masses. For this reason, our sample consists of blazars -- AGNs with their jets oriented toward Earth \cite{ulrich97} -- and GRBs, the spectral energy distributions of which are completely dominated by the jet due to beaming effects.

We used as a proxy of the jet bolometric luminosity the observed $\gamma$-ray luminosity $L^{\rm iso}$ which is isotropically equivalent. In order to estimate the kinetic power $P_{\rm jet}$, we use extended radio luminosities for the blazars whereas for the GRBs we relied on the afterglow measurements in radio or X-rays. Therefore, the availability of these observables restricted our sample to 234 blazars (106 BL Lacs and 128 flat-spectrum radio quasars -- FSRQs; see Table S1) and 54 GRBs (49 long and 5 short GRBs, all with known redshifts $z$; see Table S2). 
For blazars, $L^{\rm iso}$ was estimated from the $\gamma$-ray energy flux and the spectral index measured with Fermi Large Area Telescope (LAT) \cite{2lac}; $P_{\rm jet}$ was estimated using an empirical correlation which relates the Very Large Array (VLA) extended radio emission and the jet kinetic power \cite{cava10,meyer11}. 
For GRBs, $L^{\rm iso} = E^{\rm iso}(1+z)/t_{90}$ where $t_{90}$ is the burst duration and $E^{\rm iso}$ is the isotropically equivalent energy radiated during the prompt emission phase and measured with different telescopes (21 observed with either BeppoSAX, BATSE, HETE, HETE-2 or Integral, 24 with Swift Burst Alert Telescope -- BAT -- and 10 with Fermi). $P_{\rm jet}$ was computed as $P_{\rm jet} = f_b E_k^{\rm iso}(1+z)/t_{90}$ where $E_k^{\rm iso}$ is the kinetic energy estimated from the radio (VLA) or X-ray (Chandra) luminosity during the afterglow phase using the standard afterglow model \cite{freedman01}, $f_b \equiv 1-\cos \theta$ is the ``beaming factor'' and $\theta$ is the radiation cone half-opening angle which is the same as the jet opening angle estimated from the jet break in the GRB afterglow lightcurve \cite{online}.

\begin{figure}[!t]
\centering
\includegraphics[width=14cm,trim=80 0 100 30,clip=true]{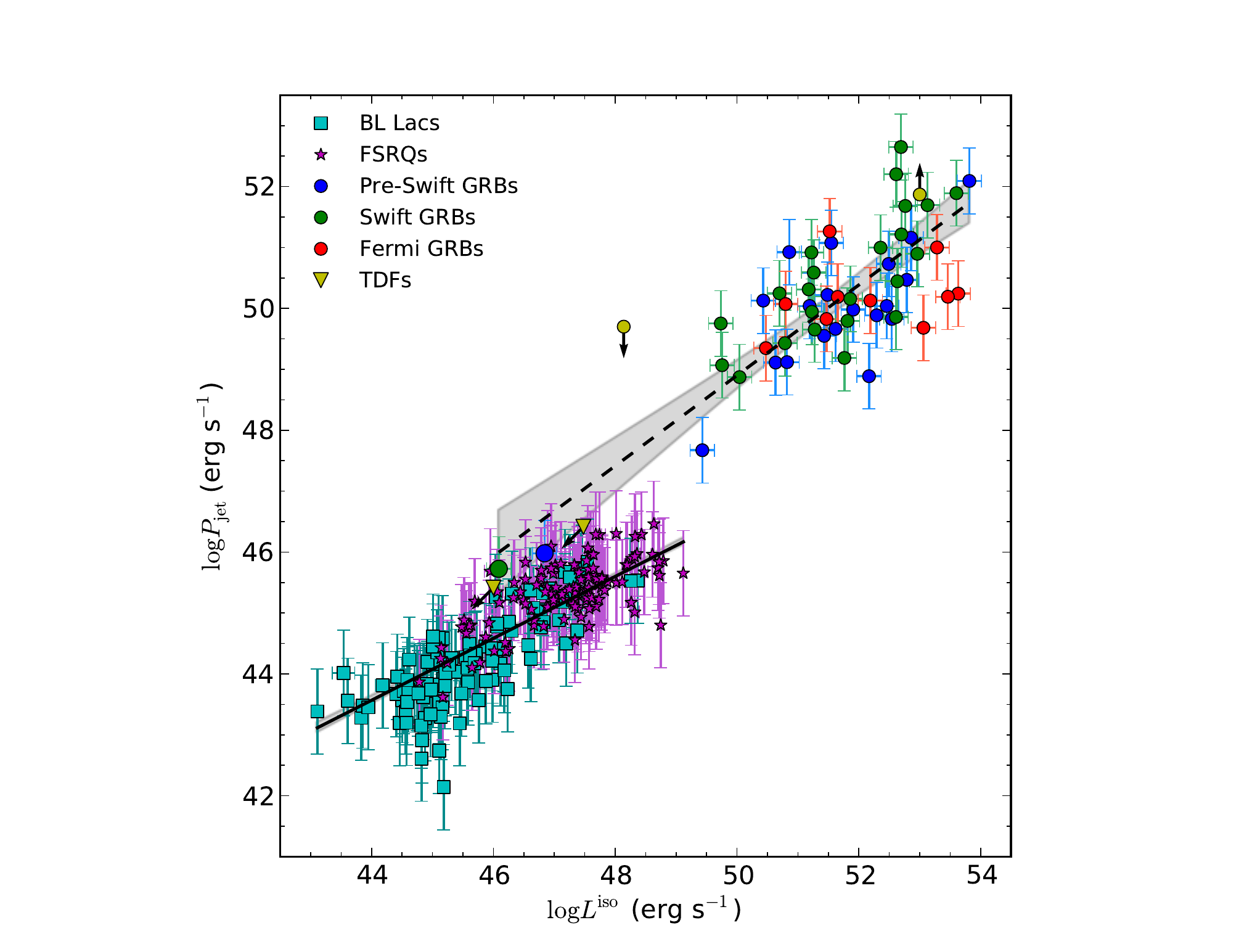}	% beam.py
\caption{The relation between the jet kinetic power and the isotropically-equivalent $\gamma$-ray luminosity for AGNs and GRBs. Error bars, $1\sigma$. 
We fitted the two populations separately using a symmetric least-squares method (orthogonal BCES with bootstrapping; \citen{bces}). The blazar and GRB best-fit models corresponds to the solid and dashed lines, respectively ($\log P_{\rm jet} = A \log L^{\rm iso} + B$). The best-fit parameters obtained for the blazars are $A=0.51 \pm 0.02$ and $B=21.2 \pm 1.1$; for the GRBs, $A=0.74 \pm 0.08$ and $B=11.8 \pm 4.1$. The scatter about the best-fit is 0.5 dex and 0.8 dex for the blazars and GRBs, respectively.
The $2\sigma$ confidence band of the fits is shown as the gray shaded regions (barely visible for blazars). The two correlations do not agree at $>5\sigma$ level.
We also include for illustration XRF 020903 and GRB 090423 (yellow circles) as well as the two recent tidal disruption flares (TDFs) detected with Swift which are presumably due to the onset of relativistic jets from the tidal disruption of stars by supermassive black holes \cite{tdfs}. We do not consider these sources in the statistics since we only have limits on their luminosities}.
\label{fig:main}
\end{figure}

We first compared the relative trends of $L^{\rm iso}$ and $P_{\rm jet}$ for the blazar and GRB population separately (Figure \ref{fig:main}). The Pearson correlation coefficients of 0.85 and 0.8 obtained for blazars and GRBs respectively, indicate a strong correlation within each group of sources. However, the $L^{\rm iso}$-$P_{\rm jet}$ trend is different for GRBs and blazars as shown by the fits to the data (Fig. \ref{fig:main}). 

\begin{figure}[!t]
\centering
\includegraphics[width=14cm,trim=20 0 50 40,clip=true]{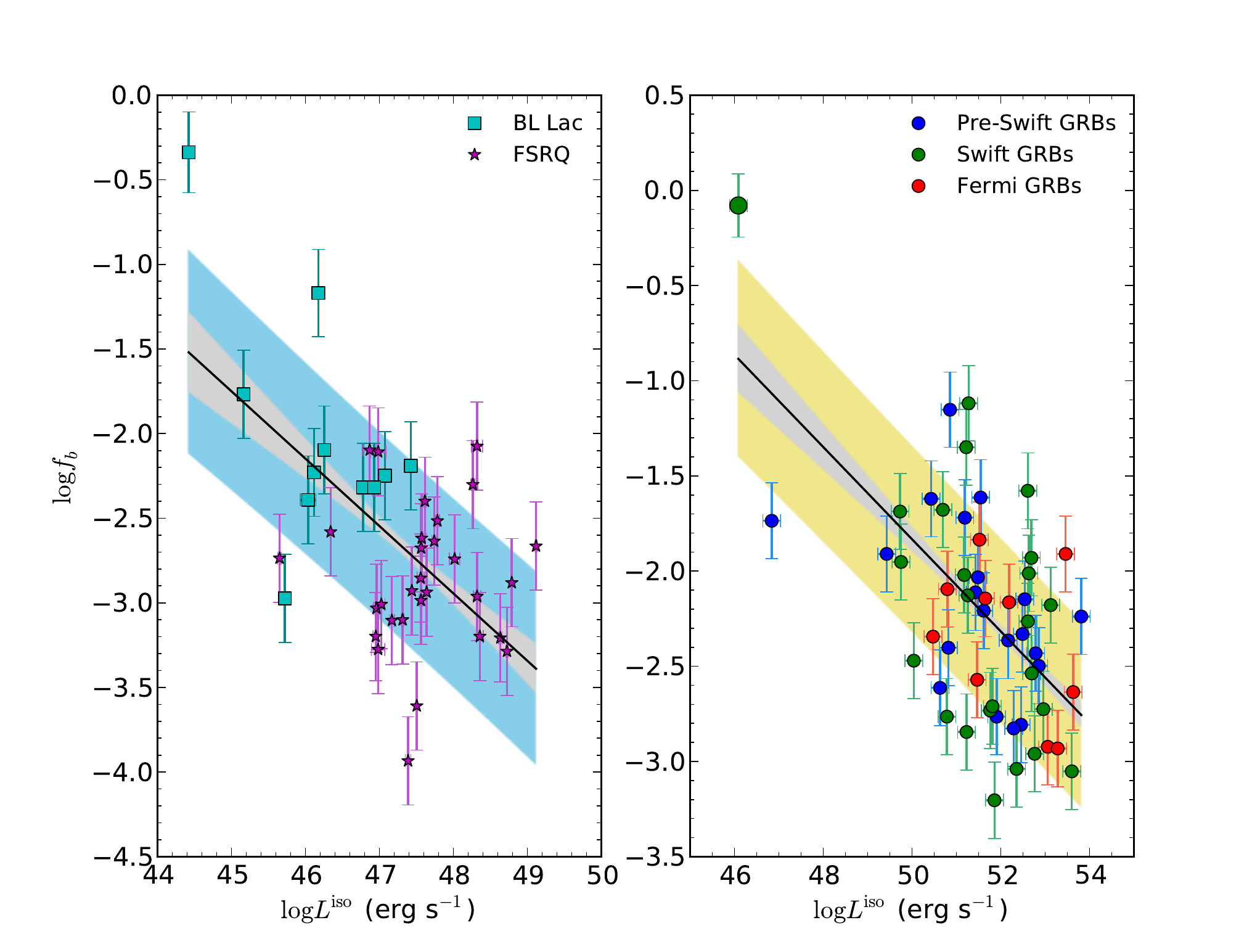}	% beaming.py
\caption{The relation between the apparent $\gamma$-ray luminosity and the beaming factor for blazars (left panel) and GRBs (right panel). 
We find $r=-0.53$ and $-0.56$ for blazars and GRBs, respectively, indicating anti correlations significant at the $3.6\sigma$ and $4.4\sigma$ levels respectively. The solid lines correspond to the best-fit linear models obtained with the symmetric least-squares fit and are given by $f_b \approx 5 \times 10^{-4} (L_{49}^{\rm iso})^{-0.39 \pm 0.15}$ and $\approx 0.03 (L_{49}^{\rm iso})^{-0.24 \pm 0.06}$ for blazars and GRBs, respectively.
The gray shaded region corresponds to the $1\sigma$ confidence band and the blue and yellow regions are the $1\sigma$ prediction bands, which quantify the scatter about the best-fits.}
\label{fig:beaming}
\end{figure}

We computed the intrinsic luminosity, $L$, for GRBs and blazars by correcting  $L^{\rm iso}$ for the opening angle or beaming factor, $f_b$, such that $L = f_b L^{\rm iso}$. For GRBs, the beaming factor is computed from the jet opening angle $\theta_j$ as $1-\cos \theta_j$ \cite{frail01}; for blazars, $f_b$ is estimated as $1-\cos 1/\Gamma$ where $\Gamma$ is the bulk Lorentz factor of the flow, since AGNs obey $\theta_{\rm j}<1/\Gamma$ \cite{jorstad05,push09}.
While an estimate of $\theta_j$ is available for each GRB in the sample, $\Gamma$ is only available for a subset of 41 blazars. Figure \ref{fig:beaming} shows an anti correlation between $L^{\rm iso}$ and $f_b$ for both GRBs and blazars with compatible indices when fit with a power law. Because $\theta$ is not available for the whole blazar sample, we used the power-law fit of $L^{\rm iso}$ vs $f_b$ as an estimator for $f_b$.

\begin{figure}[!t]
\centering
\includegraphics[width=14cm,trim=100 0 100 40,clip=true]{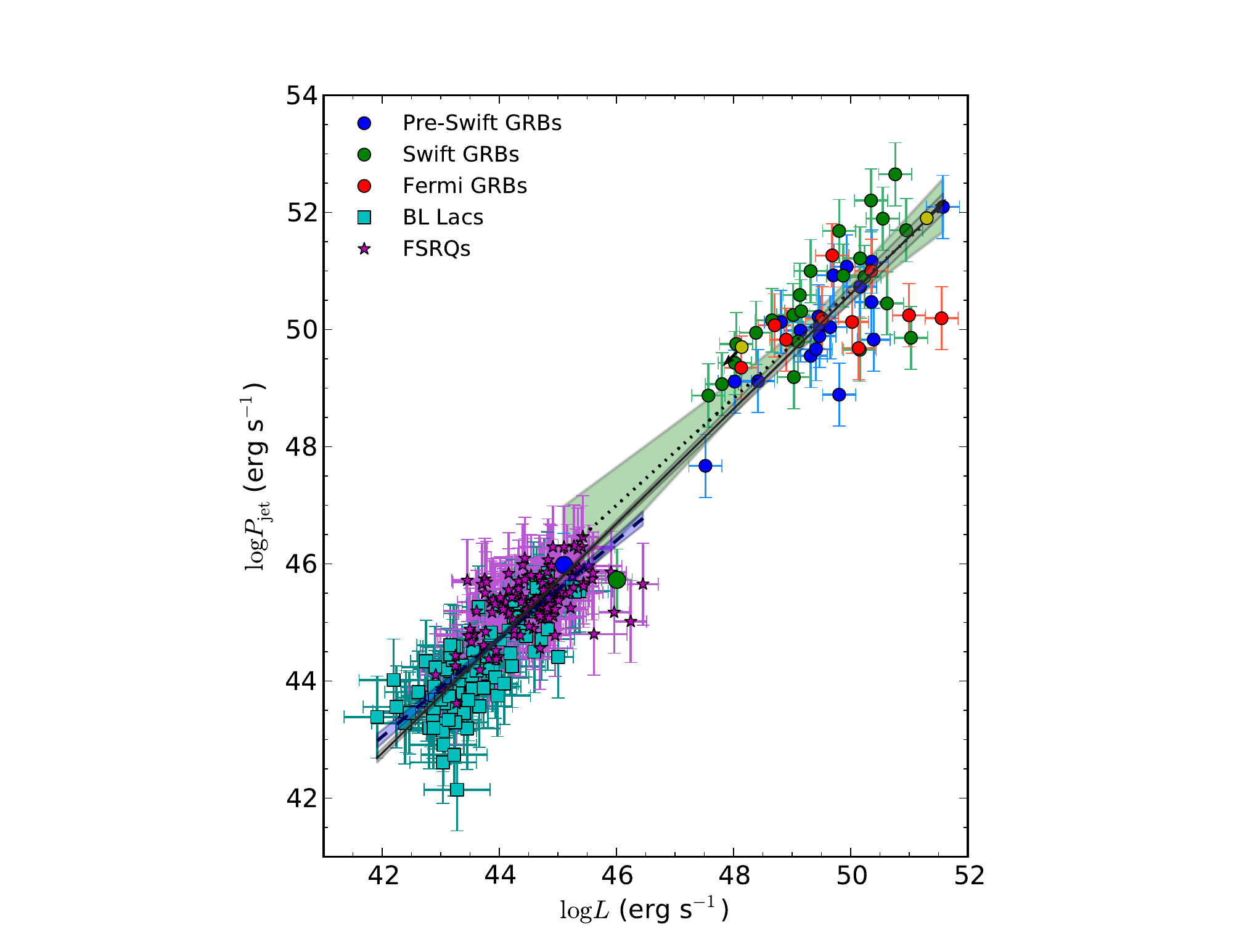}	% debeam.py
\caption{The relation between the collimation-corrected $\gamma$-ray luminosity $L=f_b L^{\rm iso}$ and the kinetic power for AGNs and GRBs. The shaded regions display the $2\sigma$ confidence band of the fits. 
The blazar and GRB best-fit models (dashed and dotted lines, respectively) follow correlations which are consistent, within the uncertainties, with the best-fit model obtained from the joint data set (solid line). In other words, using $L$ instead of $L^{\rm iso}$ leads to correlations for AGNs and GRBs which are consistent with each other (compare to Fig. \ref{fig:main}). The best-fit parameters obtained from the combined data set are $\alpha=0.98 \pm 0.02$ and $\beta=1.6 \pm 0.9$ where $\log P_{\rm jet} = \alpha \log L + \beta$. The scatter about the best-fit is 0.64 dex.  
The yellow data points correspond to XRF 020903 and GRB 090423, which we do not take into account in the statistics.}
\label{fig:debeam}
\end{figure}

As with $L^{\rm iso}$ and $P_{\rm jet}$, $L$ and $P_{\rm jet}$ are strongly correlated within the GRB and AGN samples (Fig. \ref{fig:debeam}). However, they follow the same trend within the narrow uncertainties and the whole GRB and blazar sample can be fit adequately with a power law over 10 orders of magnitude in luminosity. 
Therefore, the relativistic jets in GRBs and blazars are consistent with obeying the relation $P_{\rm jet} \approx 4.6 \times 10^{47} \left( L/10^{47} \right)^{0.98} \ {\rm erg \; s}^{-1}$, within the measurement uncertainties.
In other words, once ``black hole engines'' produce relativistic jets, they seem to do so maintaining the same coupling 
between the total power carried by the jet and power radiated away. This universal scaling for the energetics of jets is maintained across the mass scale regardless of the different environments and accretion flow conditions around the compact object. 

\begin{figure}
\centering
\includegraphics[width=14cm]{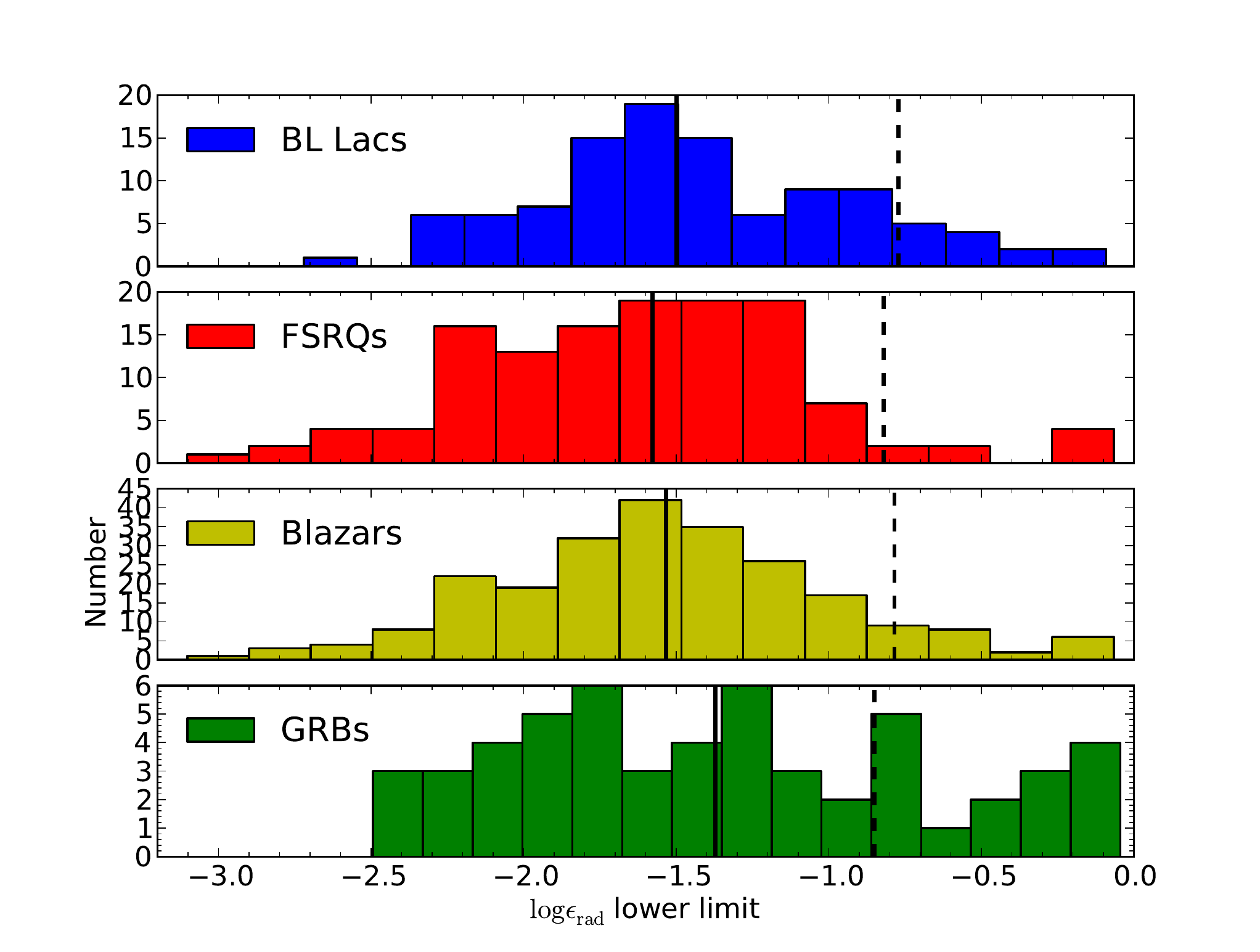}	% radeff.py
\caption{Distribution of the $1\sigma$ lower limits on the jet radiative efficiency (the fraction of the total jet power which is converted to $\gamma$-rays) $\epsilon_{\rm rad} \equiv L/(L+P_{\rm jet})$ for AGNs and GRBs. The vertical solid lines indicate the median values of the lower limits and the dashed lines represent the median values of $\epsilon_{\rm rad}$ for each sample. Most of the sources are characterized by $\epsilon_{\rm rad} > 3\%$. The median efficiencies correspond to about $15\%$, keeping in mind that these estimates are affected by $\sim 0.5-0.7$ dex uncertainties on average. }
\label{fig:eff}
\end{figure}

Figure \ref{fig:eff} indicates that most of the jets in our sample dissipate at least $3\%$ of the power carried by the jet as radiation and overall they can radiate as much $15\%$. This range of efficiencies is considerably higher than previous estimates for AGNs based on radio to X-rays luminosities  \cite{celotti93,yuan09} but they are in agreement with results obtained from blazar broadband spectral models \cite{celotti08,ghise10} as well as GRB afterglow studies \cite{fan06,zhang07,racusin11}.
Efficient heating of electrons seems to be a universal property of relativistic magnetized shocks according to numerical simulations \cite{sironi11} which demonstrate that electrons retain $\gtrsim 15\%$ of the pre-shock energy. If most of the post-shock energy is radiated away, these theoretical results could pave the way to an understanding of the high dissipation efficiencies that we find.

Our results suggest that there is a single fundamental mechanism to produce relativistic jets in the Universe. The analogy known to exist between microquasars and AGNs \cite{mirabel99,marscher02,merloni03} can be extended to the gamma-ray bursts with the fundamental difference that whereas AGNs and microquasars undergo recurrent activity, GRBs experience only one episode of hyperaccretion.

\clearpage

\bibliographystyle{Science}

\clearpage
\setcounter{page}{1}

\section*{Supplementary Materials}

\subsection*{Data}

Table S1 lists the properties of the 234 blazars in our sample (106 BL Lac objects and 128 FSRQs). Our sample includes the well-studied blazars OJ 287, 3C 454.3, 3C 279 and 3C 273. The redshifts were estimated as described in \cite{smeyer11}.

Table S2 lists the corresponding properties of the 54 GRBs (49 long and 5 short). 
We include in our data the sub-energetic bursts GRB 031203 \cite{ssode04} and GRB 980425 \cite{spain00}. The latter one is associated with the nearby (distance $\sim$ 40 Mpc) supernova 1998bw \cite{sgalama98, skulkarni98}. We also include in our sample the naked-eye GRB 080319B \cite{sracusin08}. We include in Table S2 the X-ray flash (XRF) 020903 \cite{ssode04apj} and the most distant cosmic explosion ever detected, GRB 090423 at $z \approx 8.2$ \cite{stanvir09,schandra10} but we do not take these GRBs into account in the statistics, since we have only limits on their collimation-corrected energetics.

\subsection*{Calculation of $\gamma$-ray luminosity and jet power}

\subsubsection*{Blazars}

\emph{$\gamma$-ray luminosity:}

In order to calculate the total Fermi $\gamma$-ray luminosity, we follow a procedure similar to that of \cite{sgmt09} (their equation 1). The procedure to calculate the $k$-corrected band luminosity depends on the type of model used in the 2FGL analysis \cite{s2fgl,s2lac}. In the case of a power law energy model, the total 100 MeV to 100 GeV $k$-corrected luminosity is calculated from the energy flux ($S_\gamma$) given in the catalog:
\begin{equation}
L^{\rm iso} = 4\pi d_L^2\frac{S_{\gamma}}{(1 + z)^{1-\alpha_\gamma}},
\end{equation}
where $d_L$ is the luminosity distance in cm$^2$ and $\alpha_\gamma$ is the (energy) spectral slope over the whole band. 

In the 2FGL catalog, some blazars are now modeled with the 'Log
Parabolic' form. The $k$-corrected energy flux in this case must be calculated numerically.  The integral form is
\begin{equation}
S_\gamma^\prime = \chi \int_{E_1}^{E_2} K 
\left(
  \frac{E}{E_0(1+z)}
\right)
^{ -\alpha -\beta
\mathrm{log} \left(\frac{E}{E_0(1+z)}\right)}\left(1+z\right)^{-2}E\,dE,
\end{equation}
where $E_1$ = 0.1 GeV and $E_2$ = 100 GeV and we have used the fit given in the 2FGL catalog for each source, with values $\alpha$ (column name `spectral\_index'), $\beta$ (`beta'), and $E_0$ (`pivot\_energy'), and K (`flux\_density'). The constant $\chi$ = 1.6 erg MeV GeV$^{-2}$ gives the energy flux in final units of erg cm$^{-2}$ s$^{-1}$.

The band luminosity for the Log-Parabolic model case can then be
simply calculated from
\begin{equation}
L^{\rm iso}  = 4\pi d_L^2\,S_{\gamma}^\prime.
\end{equation}

We calculate the uncertainty in $L^{\rm iso}$ propagating the error associated with $\alpha_\gamma$ and $S_\gamma$ quoted in the 2FGL. The average uncertainty in $L^{\rm iso}$ corresponds to 0.05 dex.

In order to estimate the uncertainty affecting the collimation-corrected luminosities $L$ we first evaluate the error in the beaming angle $\theta$ or correspondingly $\Gamma$.
The uncertainty in $\Gamma$ is dominated by the uncertainty in the variability Doppler factor whereas the uncertainty in the apparent speed does not contribute significantly to the error budget of $\Gamma$. The uncertainty in the variability Doppler factor is $\approx 27\%$ (1 s.d.; \emph{\citen{shovatta09}}). Therefore, for the blazars with direct estimates of $\Gamma$ available, the relative uncertainty in $\Gamma$ is $0.3$ \cite{shovatta09,spush09} which translates to an average uncertainty of 0.26 dex in $L$ for these blazars.
For the blazars without direct estimates of $\Gamma$, we estimate the uncertainty in $L$ using the prediction band of the $L^{\rm iso}-f_b$ relation shown in Fig. \ref{fig:beaming}. The plotted prediction band corresponds to a relative uncertainty of 0.69 in $\theta$. The resulting average uncertainty affecting $L$ for the blazars without direct estimates of $\Gamma$ is then 0.6 dex.

\emph{Kinetic power:}

Following \cite{smeyer11}, we estimate the jet kinetic power by using the correlation between the extended radio emission and the jet power \cite{scava10, ssullivan11}. Cavagnolo et al. searched for X-ray cavities  in different systems including giant elliptical galaxies and cD galaxies and estimated the jet power required to inflate these cavities or bubbles,  obtaining the tight correlation 
\begin{equation}	\label{eq:cava}
P_{\rm cav} \approx 6 \times 10^{43} \left( \frac{P_{\rm radio}}{{10^{40} \ \rm erg \; s}^{-1}} \right)^{0.7} {{\rm erg \; s}^{-1}}
\end{equation}
between the ``cavity'' power and the radio luminosity. Hence, assuming $P_{\rm jet} = P_{\rm cav}$ we can estimate the jet kinetic power for the blazars which have extended radio emission observed with the VLA \cite{smeyer11}.

The uncertainty in $P_{\rm jet}$ is dominated by the scatter in the correlation of \cite{scava10} and corresponds to 0.7 dex.

\subsubsection*{GRBs}

\emph{$\gamma$-ray luminosity:}

The current scenario for GRBs \cite{smeszaros02,snakar10} posits that initially most of the energy produced by the GRB is in kinetic form produced during the short ``active'' state of the stellar-mass central engine. A certain fraction of initial energy is converted after a few seconds mostly to $\gamma$-rays observed during the prompt emission, by means of internal shocks in the jet \cite{spiran99}. The ultrarelativistic jet produced in the explosion later on collides with the circumburst medium producing the afterglow. Two crucial quantities which we use in this work are the radiative and kinetic energies released by the GRBs during their short period of activity.

The isotropically equivalent energy radiated in $\gamma$-rays $E_\gamma^{\rm iso}$ is directly available from measurements. It was measured for the GRBs using a variety of different telescopes including pre-Swift telescopes (BeppoSAX, BATSE, HETE, HETE-2 and Integral) as well as Swift and Fermi (see Table S2).
We calculate the isotropically-equivalent $\gamma$-ray luminosity as
\begin{equation}
L^{\rm iso} = \frac{(1+z)}{t_{90}} E_\gamma^{\rm iso}
\end{equation}
where $t_{90}$ is the duration containing $90\%$ of the fluence in the observer frame.

The energy range in which the fluence is measured is typically $\sim 10$ keV -- 10 MeV. Most of the radiative energy released in the GRB jet during the prompt emission is contained in this energy range according to the latest GRB SEDs observed \cite{sacker11}. 

We adopt an uncertainty of 0.2 dex on the values of $E_\gamma^{\rm iso}$ which corresponds to the typical uncertainty affecting the GRBs in the sample studied by \cite{slz04}. Therefore, the resulting uncertainty in the value of $L^{\rm iso}$ corresponds to 0.2 dex.

The collimation-corrected $\gamma$-ray luminosity is computed as
\begin{equation}
L = \frac{(1+z)}{t_{90}} f_b E_\gamma^{\rm iso}
\end{equation}
where $f_b$ is the beaming factor ($f_b = 1-\cos \theta_{\rm j}$) and $\theta_{\rm j}$ is the jet half-opening angle. This relies on the afterglow lightcurve displaying a jet break which is used to estimate $\theta$ \cite{sfrail01}. $\left \langle \theta_{\rm j} \right \rangle \approx 8^\circ$ for the sample and $\left \langle f_b \right \rangle \approx 9 \times 10^{-3}$. 
We adopt an uncertainty of 0.1 dex on the values of $\theta_{\rm jet}$, which corresponds to the typical uncertainty in the values of $\theta_{\rm j}$ for the sample studied by \cite{slz04}.

We calculated the uncertainty in $L$ using error propagation from the uncertainties in $E_\gamma^{\rm iso}$ and $\theta$, obtaining that the uncertainty in $L$ is $\approx 0.3$ dex.

\emph{Kinetic energy:}

The jet kinetic energy is estimated from the radio or X-ray afterglow lightcurve using the fireball model \cite{sfreedman01}. For most of the GRBs, X-ray data were used to determine this energy \cite{slz04,sracusin11}. In a few cases, radio data were used \cite{ssode04, ssode04apj, scenko12}.

The standard fireball afterglow model depends on five model parameters: the explosion kinetic energy $E_k^{\rm iso}$, the density of the circumburst environment $n$ (with which the jet collides), the spectral index of the electron energy distribution $p$ and the fractions $\epsilon_e$ and $\epsilon_B$ of shock thermal energy carried by electrons and magnetic field, respectively. The typical values adopted for these parameters are $p=2.2$, $n=1 \ {\rm cm}^{-3}$, $\epsilon_e=0.1$ and $\epsilon_B=0.01$ (e.g., \emph{\citen{sfreedman01,sracusin11}}). 
The afterglow model relates the specific flux at a certain frequency and at a specific time after the burst (typically 10 hours in the observed frame) with the kinetic energy. 

The measurement of $E_k^{\rm iso}$ can be impacted by the afterglow plateau which possibly corresponds to a late activity of the central engine. Indeed, \cite{szhang07} demonstrated the impact of choosing two different times for the measurement of $E_k^{\rm iso}$: $t_{\rm dec}$ (deceleration time) or $t_b$ (injection break time). Adopting $t_{\rm dec}$ would lead to an underestimation of $E_k^{\rm iso}$ because this choice of time does not include the plateau.

The kinetic energy estimates we used in our analysis were computed at either $\sim 10$ h or $\sim 24$ h (cf. Table S2). We verified that these times are usually beyond the ``break time'' of the plateau reported in \cite{sdainotti10}. Therefore, the measurements of $E_k^{\rm iso}$ used in this paper correspond to conservative estimates. Moreover, the choice of the time at either $\sim 10$ h and $\sim 24$ h affects very little the measurement of $E_k^{\rm iso}$ \cite{sracusin11}.

The typical uncertainty in $E_k^{\rm iso}$ due to observational errors is $\approx 0.3$ dex \cite{slz04,szhang07}. The value of $E_k^{\rm iso}$ is also sensitive to the values of parameters which regulate the microphysics of the fireball afterglow model and are poorly constrained. The systematic error affecting $E_k^{\rm iso}$ due to the uncertainties in the parameters $\epsilon_e$ and $\epsilon_B$ of the fireball afterglow model can be as high as 0.45 dex \cite{sfan06, szhang07}. Therefore, we combined the uncertainty resulting from the observational and systematic sources of errors in quadrature and conservatively adopt an uncertainty of 0.5 dex for $E_k^{\rm iso}$.

\emph{Kinetic power:}

The jet kinetic power is computed as
\begin{equation}
P_{\rm jet} = \frac{(1+z)}{t_{90}} f_b E_k^{\rm iso}.
\end{equation}
$L_\gamma^{iso}$ and $P_{\rm jet}$ should be thought as the average luminosities over the duration $t_{90}$ of the prompt emission phase, i.e. the average luminosities over the timescale during which the central engine is producing the jet.
We calculated the uncertainty in $P_{\rm jet}$ using error propagation from the uncertainties in $E_k^{\rm iso}$ and $\theta$, obtaining that the uncertainty affecting $P_{\rm jet}$ is $\approx 0.54$ dex.

\subsection*{Linear regression method}

Here we present more details about the linear regression method that we use in the paper. 

We fitted the datasets using the BCES (bivariate correlated errors and intrinsic scatter) regression method \cite{sbces} which takes into account measurement errors in both the ``$X$'' and ``$Y$'' coordinates and the intrinsic scatter in the data. This method has been widely used in fitting datasets in the astronomical community \cite{svester06,spratt09}.

It is not clear in the data sets analyzed in this work which quantities should be treated as the dependent variables and which  ones should be treated as independent from a physical point of view. There is no a priori reason to expect the luminosity to be the independent variable as opposed to the kinetic power (or beaming factor). For this reason, we treat the variables symmetrically and adopt the BCES orthogonal regression method, which minimizes the squared orthogonal distances. Uncertainties on the parameters derived from the fits are estimated after carrying out 100000 bootstrap resamples of the data.

\subsection*{Partial correlation analysis}

When studying correlations between luminosities one should be careful to take into account their common dependence on the distance \cite{smerloni03}. We performed a partial correlation analysis of the common dependence of $L_\gamma^{\rm iso}$ and $P_{\rm jet}$ on the distance using the partial Kendall's $\tau$ correlation test \cite{spca}. 

We applied this test to our data considering $X=P_{\rm jet}$, $Y=L^{\rm iso}$ and $Z=\log d_L$. Table S3 lists the results of the partial correlation analysis.
This test demonstrates that the $p$-value of the null hypothesis (i.e. no correlation between $X$ and $Y$) is $1.2 \times 10^{-7}$ when considering the GRB subsample and $<10^{-10}$ when considering the blazar subsample or the combined blazar-GRB sample. Therefore, the $L^{\rm iso} - P_{\rm jet}$ correlation is strong and not a distance-driven artifact.

\begin{table}
\begin{center}
\begin{tabular}{cccccc}
\hline\hline
Objects & N & $\tau$ & $\sigma$ & $P_{\rm null}$ & Signif. rejection null \\
\hline
Blazars & 234 & 0.3 & 0.04 & $5 \times 10^{-15}$ & 7.8 \\
GRBs & 54 & 0.4 & 0.08 & $1.6 \times 10^{-7}$ & 5.2 \\
\hline
\end{tabular}
\label{tab:pca1}
\end{center}
{\bf Table~S3} Results of partial correlation analysis with $X=P_{\rm jet}$, $Y=L^{\rm iso}$ and $Z=\log d_L$. \\
{\bf Notes.} Column (1): subsample. Column (2): Number of sources. Column (3)-(6): results of partial correlation analysis; $\tau$ is the partial Kendall's correlation coefficient; $\sigma$ is the square root of the calculated variance; $P_{\rm null}$ is the probability for accepting the null hypothesis that there is no correlation between $X$ and $Y$; Column (6) gives the associated significance in standard deviations with which the null hypothesis is rejected.
\end{table}

We then applied the partial correlation test to the data considering $X=P_{\rm jet}$, $Y=L$ and $Z=\log d_L$, i.e. we replaced the isotropically-equivalent luminosity in $Y$ with the collimation-corrected one. Table S4 lists the results of the analysis. 
This test demonstrates that the p-value of the null hypothesis is $\ll 10^{-10}$, $\approx 10^{-12}$ and $10^{-9}$ when considering the combined blazar-GRB sample, the blazars and the GRB, respectively. Hence, the $L - P_{\rm jet}$ correlation remains very strong after correcting for beaming. 

\begin{table}
\begin{center}
\begin{tabular}{cccccc}
\hline\hline
Objects & N & $\tau$ & $\sigma$ & $P_{\rm null}$ & Signif. rejection null \\
\hline
All objects	 & 288 & 0.57 & 0.06 & $\ll 10^{-10}$ & 9.4 \\
Blazars & 234 & 0.24 & 0.03 & $2 \times 10^{-12}$ & 7 \\
GRBs & 54 & 0.47 & 0.08 & $10^{-9}$ & 6.1 \\
\hline
\end{tabular}
\label{tab:pca2}
\end{center}
{\bf Table~S4} Same as Table S3 with $X=P_{\rm jet}$, $Y=L$ and $Z=\log d_L$.
\end{table}

\subsection*{Blazar luminosity estimates: Impact of the synchrotron peak}

We discuss in this section the impact of the lower energy synchrotron peak -- observed in the spectral energy distribution of blazars \cite{sfossati98,sdonato01} -- in estimating the jet radiative luminosity. 

A subset of 131 blazars -- roughly half of the original AGN sample -- have adequate sampling of their multiwavelength spectral energy distributions which allow us to quantify the impact of the synchrotron peak in the estimate of the jet radiative luminosity. We estimated the bolometric luminosity as $L_{\rm bol}^{\rm iso}=L^{\rm iso}+L_{\rm syn}^{\rm iso}$ where $L_{\rm syn}^{\rm iso}$ is the isotropically-equivalent luminosity of the synchrotron peak. We obtained that the values of $L_{\rm bol}^{\rm iso}$ for the quasars are not much different from $L^{\rm iso}$ (on average by a factor of $\approx 1.7$), whereas the values of $L_{\rm bol}^{\rm iso}$ for the BL Lacs are somewhat different (on average by a factor of $\approx 2.8$).

We computed the intrinsic bolometric luminosity as $L_{\rm bol}=f_b L_{\rm bol}^{\rm iso}$,  correcting $L_{\rm bol}^{\rm iso}$ for the opening angle or beaming factor. As previously discussed, estimates of $f_b$ for blazars rely on measurements of $\Gamma$ which are not available for all the blazars in our sample. For this reason, we use the anti correlation between $L_{\rm bol}^{\rm iso}$ and $f_b$ as an estimator of $f_b$ for the blazars without measurements of $\Gamma$.

We show in Figure S1 the resulting relation between $L_{\rm bol}$ and $P_{\rm jet}$ compared to the $L - P_{\rm jet}$ fit derived before (cf. Fig. \ref{fig:debeam}). Figure S1 illustrates that the $L_{\rm bol} - P_{\rm jet}$ and $L - P_{\rm jet}$ best-fits are characterized by very similar slopes. However, the fit based on $L_{\rm bol}$ is characterized by slightly higher radiative efficiencies: for a given jet power, the jet luminosity is higher on average by a factor of $\approx 2$ compared to the fit based on $L$. 
Therefore, the results based on $L_{\rm bol}$ strengthen our conclusion that AGN jets have high radiative efficiencies while still being in qualitative agreement with the GRB result.

\begin{figure}[!t]
\centering
\includegraphics[width=14cm,trim=100 0 100 40,clip=true]{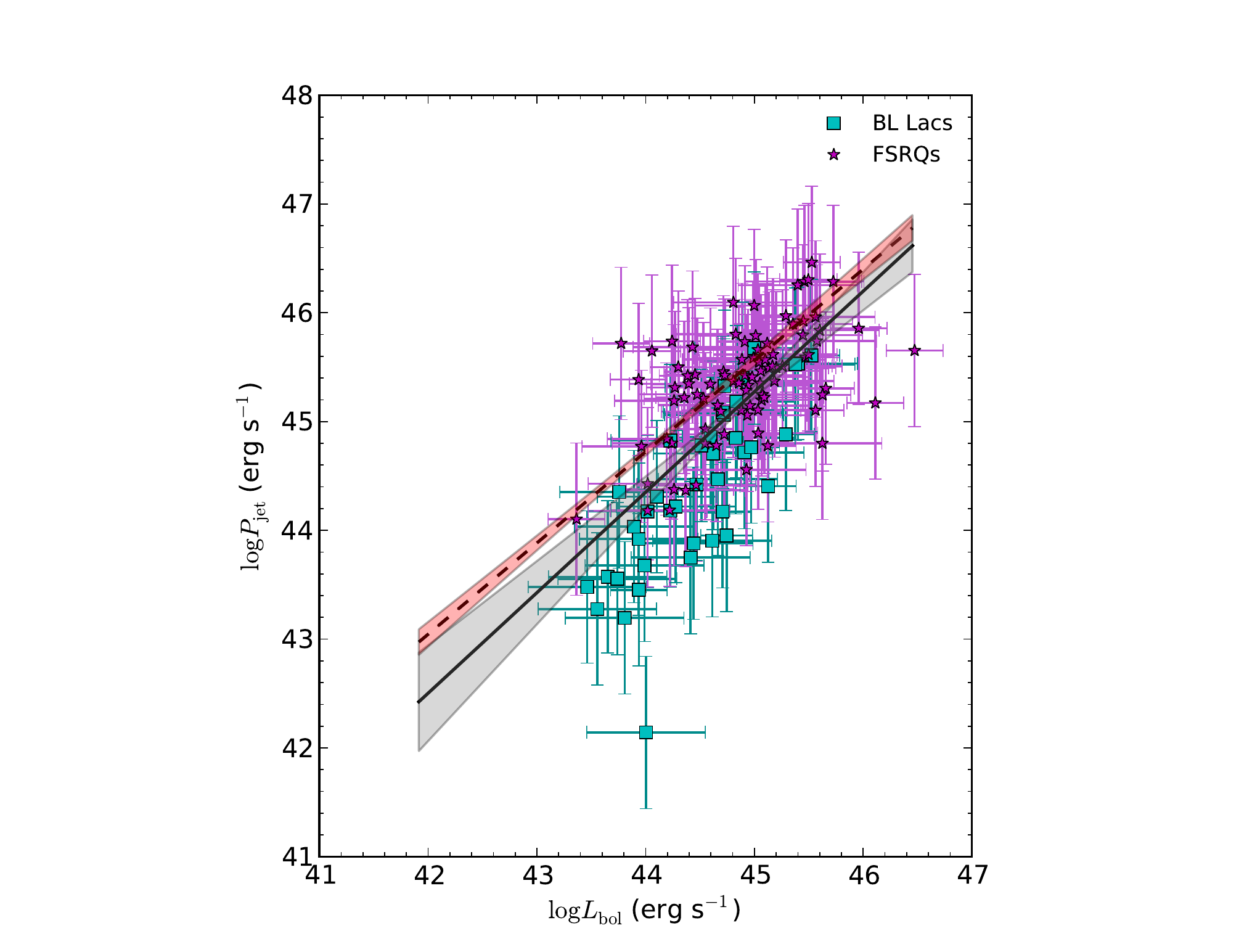}	% misc/bolcorr/compare.py
\\
{\bf Figure~S1} The relation between the collimation-corrected bolometric luminosity $L_{\rm bol}$ and the kinetic power for the 131 blazars with measurements of synchrotron peak luminosity. The solid line in this figure corresponds to the fit based on $L_{\rm bol}$ whereas the dashed line is the fit based on $L$. The shaded regions display the $2\sigma$ confidence band of the fits. Error bars, $1 \sigma$.
\label{fig:compare}
\end{figure}

\clearpage

\begin{landscape}
\begin{longtable}{cccccccccc}
\multicolumn{10}{l}{{\bf Table~S1.} Properties of the blazar sample. } \\
\\
\hline\hline
2FGL name & Alias & Type & $z$ & $\log L^{\rm iso}$ & Unc.$^{\rm a}$ & $\log L$ & Unc.$^{\rm b}$ & $\log P_{\rm jet}$$^{\rm c}$ & $\log f_b$$^{\rm d}$ \\
 &  &  &  & (erg s$^{-1}$) & $\log L^{\rm iso}$ & (erg s$^{-1}$) & $\log L$ & (erg s$^{-1}$) & \\
 \hline
J0757.1+0957 & PKS 0754+100 & BLL & 0.266 & 45.72 & 0.04 & 42.75 & 0.26 & 44.34 & -2.97* \\
J0811.4+0149 & PKS 0808+019 & BLL & 1.148 & 47.13 & 0.05 & 44.56 & 0.57 & 45.18 & -2.57 \\
J0807.1-0543 & PKS 0804-055 & BLL & 0.158 & 45.01 & 0.06 & 43.16 & 0.57 & 44.16 & -1.84 \\
J0808.2-0750 & PKS 0805-07 & FSRQ & 1.837 & 48.61 & 0.02 & 45.53 & 0.56 & 45.96 & -3.08 \\
J0825.9+0308 & PKS 0823+033 & BLL & 0.506 & 45.86 & 0.09 & 43.73 & 0.57 & 44.45 & -2.14 \\
J0831.9+0429 & PKS 0829+046 & BLL & 0.174 & 45.68 & 0.02 & 43.6 & 0.56 & 44.18 & -2.08 \\
J0839.4+1802 & BZB J0839+1802 & BLL & 0.28 & 45.17 & 0.1 & 43.27 & 0.57 & 44.58 & -1.9 \\
J0847.2+1134 & BZB J0847+1133 & BLL & 0.199 & 44.98 & 0.12 & 43.14 & 0.58 & 43.76 & -1.83 \\
J0839.6+0059 & PKS 0837+012 & FSRQ & 1.123 & 46.87 & 0.08 & 44.39 & 0.57 & 45.41 & -2.49 \\
J0303.5+4713 & 4C +47.08 & BLL & 0.475 & 46.3 & 0.04 & 44.01 & 0.57 & 44.71 & -2.29 \\
J0854.8+2005 & OJ 287 & BLL & 0.306 & 46.12 & 0.03 & 43.89 & 0.26 & 44.17 & -2.23* \\
J2151.5-3021 & PKS 2149-307 & FSRQ & 2.345 & 48.48 & 0.06 & 45.44 & 0.57 & 45.67 & -3.04 \\
J2158.8-3013 & PKS 2155-304 & BLL & 0.117 & 45.98 & 0.01 & 43.8 & 0.56 & 43.9 & -2.18 \\
J2258.0-2759 & PKS 2255-282 & FSRQ & 0.927 & 47.35 & 0.03 & 44.7 & 0.57 & 45.35 & -2.65 \\
J0120.4-2700 & 0118-272 & BLL & 0.557 & 46.78 & 0.03 & 44.32 & 0.57 & 45.06 & -2.45 \\
J0252.7-2218 & PKS 0250-225 & FSRQ & 1.427 & 48.02 & 0.03 & 45.14 & 0.57 & 45.5 & -2.88 \\
J2213.1-2527 & PKS 2210-25 & FSRQ & 1.831 & 47.6 & 0.08 & 44.86 & 0.57 & 45.97 & -2.73 \\
J2243.2-2540 & PKS 2240-260 & BLL & 0.774 & 46.81 & 0.04 & 44.35 & 0.57 & 45.33 & -2.47 \\
J0137.6-2430 & PKS 0135-247 & FSRQ & 0.831 & 46.78 & 0.05 & 44.33 & 0.57 & 45.57 & -2.45 \\
J0205.3-1657 & PKS 0202-17 & FSRQ & 1.74 & 47.77 & 0.06 & 44.98 & 0.57 & 45.42 & -2.79 \\
J2157.9-1501 & PKS 2155-152 & FSRQ & 0.672 & 46.44 & 0.06 & 44.11 & 0.57 & 45.34 & -2.34 \\
J0132.8-1654 & PKS 0130-17 & FSRQ & 1.02 & 47.35 & 0.03 & 44.7 & 0.57 & 45.22 & -2.65 \\
J2347.9-1629 & PKS 2345-16 & FSRQ & 0.576 & 46.5 & 0.05 & 44.14 & 0.57 & 45.21 & -2.36 \\
J0116.0-1134 & PKS 0113-118 & FSRQ & 0.672 & 46.71 & 0.03 & 44.28 & 0.57 & 45.46 & -2.43 \\
J2229.7-0832 & PKS 2227-08 & FSRQ & 1.56 & 48.26 & 0.03 & 45.96 & 0.26 & 45.17 & -2.3* \\
J0102.7+5827 & TEX 0059+581 & FSRQ & 0.643 & 46.82 & 0.03 & 44.36 & 0.57 & 44.78 & -2.47 \\
J0050.6-0929 & PKS 0048-09 & BLL & 0.2 & 45.7 & 0.03 & 43.62 & 0.56 & 44.31 & -2.08 \\
J0141.5-0928 & 0138-097 & BLL & 0.733 & 46.65 & 0.05 & 44.24 & 0.57 & 45.08 & -2.41 \\
J2225.6-0454 & 3C 446 & FSRQ & 1.404 & 47.74 & 0.03 & 45.11 & 0.26 & 46.29 & -2.64* \\
J2133.8-0154 & 4C -02.81 & BLL & 1.285 & 47.18 & 0.07 & 44.59 & 0.57 & 45.67 & -2.59 \\
J2338.1-0229 & PKS 2335-027 & FSRQ & 1.072 & 47.32 & 0.04 & 44.68 & 0.57 & 45.22 & -2.64 \\
J0108.6+0135 & PKS 0106+01 & FSRQ & 2.099 & 48.64 & 0.02 & 45.43 & 0.26 & 46.46 & -3.21* \\
J0217.9+0143 & PKS 0215+015 & FSRQ & 1.715 & 48.18 & 0.02 & 45.24 & 0.56 & 45.79 & -2.93 \\
J2148.2+0659 & 4C +06.69 & FSRQ & 0.99 & 46.87 & 0.09 & 44.77 & 0.28 & 45.1 & -2.1* \\
J2334.3+0734 & BZQ J2334+0736 & FSRQ & 0.401 & 45.87 & 0.06 & 43.73 & 0.57 & 44.6 & -2.14 \\
J2330.2+1107 & 4C +10.73 & FSRQ & 1.489 & 47.34 & 0.12 & 44.7 & 0.58 & 45.4 & -2.65 \\
J0259.5+0740 & 0256+075 & FSRQ & 0.893 & 46.62 & 0.09 & 44.22 & 0.57 & 45.06 & -2.4 \\
J2147.3+0930 & 2144+092 & FSRQ & 1.113 & 47.72 & 0.02 & 44.95 & 0.56 & 45.46 & -2.78 \\
J2232.4+1143 & 4C 11.69 & FSRQ & 1.037 & 47.56 & 0.03 & 44.89 & 0.26 & 45.72 & -2.68* \\
J2253.9+1609 & 3C 454.3 & FSRQ & 0.859 & 48.79 & 0.0 & 45.91 & 0.26 & 45.86 & -2.88* \\
J2143.5+1743 & S3 2141+17 & FSRQ & 0.211 & 46.01 & 0.01 & 43.82 & 0.56 & 44.38 & -2.19 \\
J2203.4+1726 & PKS 2201+171 & FSRQ & 1.075 & 47.74 & 0.02 & 44.95 & 0.56 & 45.49 & -2.78 \\
J0238.7+1637 & AO 0235+164 & BLL & 0.524 & 47.38 & 0.01 & 44.72 & 0.56 & 44.72 & -2.66 \\
J0112.1+2245 & S2 0109+22 & BLL & 0.265 & 46.24 & 0.02 & 43.97 & 0.56 & 43.75 & -2.27 \\
J2217.1+2422 & B2 2214+24B & BLL & 0.505 & 46.06 & 0.06 & 43.85 & 0.57 & 44.78 & -2.21 \\
J2311.0+3425 & B2 2308+34 & FSRQ & 1.817 & 48.2 & 0.03 & 45.26 & 0.57 & 45.79 & -2.94 \\
J0205.4+3211 & 0202+319 & FSRQ & 1.466 & 47.34 & 0.09 & 44.69 & 0.57 & 45.33 & -2.65 \\
J0237.8+2846 & 4C +28.07 & FSRQ & 1.207 & 47.78 & 0.03 & 45.27 & 0.26 & 45.59 & -2.52* \\
J0221.0+3555 & 0218+357 & FSRQ & 0.96 & 47.56 & 0.02 & 44.84 & 0.56 & 45.58 & -2.72 \\
J0909.1+0121 & 4C +01.24 & FSRQ & 1.024 & 47.6 & 0.04 & 44.86 & 0.57 & 45.25 & -2.74 \\
J0915.8+2932 & B2 0912+29 & BLL & 0.096 & 44.75 & 0.05 & 42.99 & 0.56 & 43.76 & -1.76 \\
J0927.9-2041 & PKS 0925-203 & FSRQ & 0.348 & 45.63 & 0.07 & 43.57 & 0.57 & 44.8 & -2.06 \\
J0956.9+2516 & OK 290 & FSRQ & 0.712 & 46.67 & 0.04 & 44.25 & 0.57 & 44.8 & -2.41 \\
J1014.1+2306 & 4C +23.24 & FSRQ & 0.565 & 46.07 & 0.09 & 43.86 & 0.57 & 45.35 & -2.21 \\
J1012.1+0631 & BZB J1012+0630 & BLL & 0.727 & 46.3 & 0.09 & 44.01 & 0.57 & 45.32 & -2.29 \\
J1023.6+3947 & B2 1020+40 & FSRQ & 1.254 & 47.02 & 0.08 & 44.48 & 0.57 & 45.78 & -2.54 \\
J1051.3+3938 & BZB J1051+3943 & BLL & 0.49 & 45.57 & 0.17 & 43.53 & 0.59 & 43.93 & -2.04 \\
J1040.7+0614 & 4C +06.41 & FSRQ & 1.27 & 47.38 & 0.05 & 44.72 & 0.57 & 45.06 & -2.66 \\
J1058.4+0133 & 4C +01.28 & BLL & 0.888 & 47.42 & 0.02 & 45.23 & 0.26 & 45.61 & -2.19* \\
J1104.4+3812 & MKN 421 & BLL & 0.03 & 44.88 & 0.01 & 43.08 & 0.56 & 43.28 & -1.8 \\
J1121.0+4211 & EXO 1118.0+4228 & BLL & 0.124 & 44.82 & 0.07 & 43.04 & 0.57 & 42.61 & -1.78 \\
J1132.9+0033 & PKS 1130+008 & BLL & 1.223 & 47.32 & 0.05 & 44.68 & 0.57 & 45.46 & -2.64 \\
J1130.3-1448 & PKS 1127-145 & FSRQ & 1.187 & 47.68 & 0.03 & 44.92 & 0.57 & 45.62 & -2.76 \\
J1126.6-1856 & PKS 1124-186 & FSRQ & 1.048 & 47.52 & 0.02 & 44.81 & 0.56 & 45.2 & -2.71 \\
J1146.9+4000 & S4 1144+402 & FSRQ & 1.088 & 47.34 & 0.03 & 44.69 & 0.57 & 44.56 & -2.65 \\
J1146.8-3812 & PKS 1144-379 & FSRQ & 1.048 & 47.16 & 0.05 & 44.58 & 0.57 & 44.89 & -2.58 \\
J1159.5+2914 & 4C +29.45 & FSRQ & 0.729 & 47.31 & 0.02 & 44.21 & 0.26 & 45.43 & -3.1* \\
J1209.6+4121 & B3 1206+416 & BLL & 0.194 & 44.81 & 0.13 & 43.04 & 0.58 & 43.15 & -1.78 \\
J1206.0-2638 & PKS 1203-26 & FSRQ & 0.786 & 46.79 & 0.06 & 44.33 & 0.57 & 45.7 & -2.46 \\
J1221.3+3010 & PG 1218+304 & BLL & 0.2 & 45.61 & 0.05 & 43.56 & 0.57 & 44.49 & -2.05 \\
J1217.8+3006 & ON 325 & BLL & 0.1 & 45.18 & 0.02 & 43.28 & 0.56 & 43.92 & -1.9 \\
J1221.4+2814 & ON 231 & BLL & 0.1 & 45.18 & 0.02 & 43.28 & 0.56 & 42.14 & -1.91 \\
J1222.4+0413 & PKS 1219+04 & FSRQ & 0.965 & 47.36 & 0.04 & 44.7 & 0.57 & 45.42 & -2.65 \\
J1219.7+0201 & PKS 1217+02 & FSRQ & 0.24 & 45.14 & 0.11 & 43.25 & 0.57 & 44.43 & -1.89 \\
J1225.0+4335 & B3 1222+438 & BLL & 1.87 & 47.46 & 0.1 & 44.77 & 0.57 & 45.74 & -2.69 \\
J1231.7+2848 & B2 1229+29 & BLL & 1.0 & 47.3 & 0.04 & 44.67 & 0.57 & 45.22 & -2.63 \\
J1224.9+2122 & 4C +21.35 & FSRQ & 0.435 & 47.5 & 0.01 & 43.9 & 0.26 & 45.38 & -3.61* \\
J1229.1+0202 & 3C 273 & FSRQ & 0.158 & 46.34 & 0.01 & 43.76 & 0.26 & 45.5 & -2.58* \\
J1243.1+3627 & Ton 116 & BLL & 0.112 & 44.89 & 0.06 & 43.09 & 0.57 & 43.72 & -1.81 \\
J1256.1-0547 & 3C 279 & FSRQ & 0.536 & 47.64 & 0.01 & 44.7 & 0.26 & 45.73 & -2.94* \\
J1246.7-2546 & PKS 1244-255 & FSRQ & 0.638 & 47.31 & 0.02 & 44.67 & 0.56 & 45.14 & -2.64 \\
J1309.4+4304 & B3 1307+433 & BLL & 0.69 & 46.61 & 0.05 & 44.22 & 0.57 & 44.25 & -2.4 \\
J1310.6+3222 & 1308+326 & FSRQ & 0.997 & 47.56 & 0.03 & 44.58 & 0.26 & 45.37 & -2.99* \\
J1326.8+2210 & B2 1324+22 & FSRQ & 1.4 & 47.69 & 0.04 & 44.92 & 0.57 & 45.11 & -2.77 \\
J1332.0-0508 & PKS 1329-049 & FSRQ & 2.15 & 48.7 & 0.03 & 45.59 & 0.57 & 45.74 & -3.11 \\
J1337.7-1257 & PKS 1335-127 & FSRQ & 0.539 & 46.53 & 0.04 & 44.16 & 0.57 & 45.15 & -2.37 \\
J1354.7-1047 & PKS 1352-104 & FSRQ & 0.332 & 45.93 & 0.04 & 43.77 & 0.57 & 44.84 & -2.16 \\
J1405.1+0405 & PKS 1402+044 & FSRQ & 3.215 & 47.68 & 0.16 & 44.92 & 0.59 & 46.29 & -2.76 \\
J1419.4+3820 & B3 1417+385 & FSRQ & 1.82 & 47.58 & 0.09 & 44.85 & 0.57 & 45.07 & -2.73 \\
J1418.1+2539 & BZB J1417+2543 & BLL & 0.237 & 44.92 & 0.12 & 43.1 & 0.57 & 44.2 & -1.81 \\
J1427.0+2347 & PG 1424+240 & BLL & 0.16 & 45.98 & 0.02 & 43.8 & 0.56 & 44.22 & -2.18 \\
J1440.3-1540 & PKS 1437-153 & BLL & 0.244 & 45.1 & 0.1 & 43.22 & 0.57 & 44.59 & -1.88 \\
J1513.6-3233 & CRATES J1513-3234 & FSRQ & 0.244 & 45.6 & 0.04 & 43.55 & 0.57 & 44.79 & -2.05 \\
J1517.7-2421 & AP Lib & BLL & 0.049 & 44.5 & 0.03 & 42.83 & 0.56 & 43.57 & -1.67 \\
J0136.9+4751 & DA 55 & FSRQ & 0.859 & 47.57 & 0.02 & 44.95 & 0.26 & 44.78 & -2.62* \\
J1512.8-0906 & PKS 1510-089 & FSRQ & 0.36 & 47.44 & 0.01 & 44.51 & 0.26 & 44.93 & -2.93* \\
J1733.1-1307 & NRAO 530 & FSRQ & 0.902 & 47.39 & 0.04 & 43.45 & 0.26 & 45.72 & -3.93* \\
J2025.6-0736 & PKS 2022-077 & FSRQ & 1.388 & 48.27 & 0.02 & 45.31 & 0.56 & 45.9 & -2.97 \\
J1510.9-0545 & 4C -05.64 & FSRQ & 1.191 & 47.55 & 0.05 & 44.83 & 0.57 & 46.07 & -2.72 \\
J1512.2+0201 & PKS 1509+022 & FSRQ & 0.222 & 45.48 & 0.04 & 43.47 & 0.57 & 44.77 & -2.01 \\
J1549.5+0237 & PKS 1546+027 & FSRQ & 0.414 & 46.26 & 0.03 & 43.98 & 0.57 & 44.42 & -2.27 \\
J1728.2+0429 & PKS 1725+044 & FSRQ & 0.293 & 45.78 & 0.05 & 43.67 & 0.57 & 44.18 & -2.11 \\
J0217.7+7353 & S5 0212+73 & FSRQ & 2.367 & 48.32 & 0.07 & 46.25 & 0.27 & 45.01 & -2.07* \\
J1550.7+0526 & 4C +05.64 & FSRQ & 1.422 & 47.33 & 0.07 & 44.69 & 0.57 & 45.79 & -2.64 \\
J1751.5+0938 & OT 081 & BLL & 0.322 & 46.26 & 0.03 & 44.16 & 0.26 & 44.86 & -2.1* \\
J1608.5+1029 & 4C +10.45 & FSRQ & 1.226 & 47.56 & 0.05 & 44.7 & 0.26 & 45.29 & -2.85* \\
J2035.4+1058 & PKS 2032+107 & FSRQ & 0.601 & 46.67 & 0.04 & 44.25 & 0.57 & 44.88 & -2.42 \\
J1504.3+1029 & PKS 1502+106 & FSRQ & 1.839 & 49.12 & 0.01 & 46.45 & 0.26 & 45.65 & -2.66* \\
J1553.5+1255 & PKS 1551+130 & FSRQ & 1.29 & 47.78 & 0.03 & 44.98 & 0.57 & 45.52 & -2.8 \\
J0222.6+4302 & 3C 66A & BLL & 0.444 & 47.31 & 0.02 & 44.67 & 0.56 & 45.43 & -2.64 \\
J1838.7+4759 & BZB J1838+4802 & BLL & 0.131 & 44.83 & 0.07 & 43.04 & 0.57 & 42.91 & -1.78 \\
J1540.4+1438 & 4C +14.60 & BLL & 0.605 & 46.04 & 0.1 & 43.65 & 0.28 & 45.26 & -2.39* \\
J1607.0+1552 & PKS 1604+159 & BLL & 0.357 & 46.06 & 0.03 & 43.86 & 0.57 & 44.82 & -2.21 \\
J1719.3+1744 & PKS 1717+177 & BLL & 0.137 & 45.11 & 0.05 & 43.23 & 0.57 & 43.55 & -1.88 \\
J2009.7+7225 & 4C +72.28 & BLL & 0.156 & 45.08 & 0.04 & 43.21 & 0.57 & 44.35 & -1.87 \\
J1516.9+1925 & PKS 1514+197 & BLL & 0.65 & 46.18 & 0.09 & 43.93 & 0.57 & 44.06 & -2.25 \\
J1746.0+2316 & BZQ J1745+2252 & FSRQ & 1.884 & 47.73 & 0.11 & 44.95 & 0.58 & 45.53 & -2.78 \\
J1813.5+3143 & B2 1811+31 & BLL & 0.117 & 44.85 & 0.04 & 43.06 & 0.56 & 43.62 & -1.79 \\
J1613.4+3409 & OS 319 & FSRQ & 1.401 & 46.98 & 0.1 & 44.88 & 0.28 & 45.3 & -2.11* \\
J2004.5+7754 & S5 2007+777 & BLL & 0.342 & 45.71 & 0.05 & 43.62 & 0.57 & 44.33 & -2.09 \\
J1640.7+3945 & NRAO 512 & FSRQ & 1.66 & 48.12 & 0.04 & 45.2 & 0.57 & 45.51 & -2.91 \\
J1635.2+3810 & 4C +38.41 & FSRQ & 1.814 & 48.72 & 0.02 & 45.44 & 0.26 & 45.62 & -3.29* \\
J1642.9+3949 & 3C 345 & FSRQ & 0.593 & 46.95 & 0.04 & 43.75 & 0.26 & 45.74 & -3.2* \\
J1653.9+3945 & MKN 501 & BLL & 0.034 & 44.46 & 0.02 & 42.8 & 0.56 & 43.2 & -1.66 \\
J1749.1+4323 & B3 1747+433 & BLL & 0.229 & 45.38 & 0.05 & 43.41 & 0.57 & 44.03 & -1.97 \\
J1801.7+4405 & 1800+440 & FSRQ & 0.663 & 46.33 & 0.08 & 44.03 & 0.57 & 45.25 & -2.3 \\
J0533.0+4823 & TEX 0529+483 & FSRQ & 1.162 & 47.53 & 0.03 & 44.82 & 0.57 & 45.25 & -2.71 \\
J0456.1-4613 & PKS 0454-46 & FSRQ & 0.858 & 47.11 & 0.03 & 44.54 & 0.57 & 45.8 & -2.57 \\
J0449.4-4350 & PKS 0447-439 & BLL & 0.107 & 45.58 & 0.02 & 43.54 & 0.56 & 43.86 & -2.04 \\
J1829.2+5402 & BZB J1829+5402 & BLL & 0.181 & 44.85 & 0.08 & 43.06 & 0.57 & 43.81 & -1.79 \\
J1824.0+5650 & 4C +56.27 & BLL & 0.663 & 46.93 & 0.02 & 44.61 & 0.26 & 45.4 & -2.32* \\
J1806.7+6948 & 3C 371 & BLL & 0.051 & 44.42 & 0.02 & 44.08 & 0.24 & 43.95 & -0.34* \\
J1800.5+7829 & S5 1803+784 & BLL & 0.684 & 47.08 & 0.02 & 44.83 & 0.26 & 44.88 & -2.25* \\
J0635.5-7516 & PKS 0637-75 & FSRQ & 0.651 & 46.89 & 0.03 & 44.4 & 0.56 & 45.98 & -2.49 \\
J1118.1-4629 & PKS 1116-46 & FSRQ & 0.713 & 46.53 & 0.07 & 44.16 & 0.57 & 45.83 & -2.37 \\
J1742.1+5948 & RGB 1742+597 & BLL & 0.148 & 44.56 & 0.08 & 42.87 & 0.57 & 43.38 & -1.69 \\
J2202.8+4216 & BL Lac & BLL & 0.069 & 45.16 & 0.01 & 43.4 & 0.26 & 43.45 & -1.77* \\
J1748.8+7006 & 1749+701 & BLL & 0.77 & 46.82 & 0.04 & 44.36 & 0.57 & 44.85 & -2.47 \\
J2258.8-5524 & BZB J2258-5525 & BLL & 0.479 & 45.54 & 0.15 & 43.52 & 0.58 & 44.27 & -2.03 \\
J1740.2+5212 & 4C +51.37 & FSRQ & 1.375 & 47.83 & 0.03 & 45.01 & 0.57 & 45.37 & -2.81 \\
J1725.2+5853 & BZB J1725+5851 & BLL & 0.125 & 44.41 & 0.08 & 42.77 & 0.57 & 43.67 & -1.64 \\
J1538.1+8159 & 1ES 1544+820 & BLL & 0.138 & 44.53 & 0.12 & 42.85 & 0.58 & 43.72 & -1.68 \\
J1428.0-4206 & PKS 1424-41 & FSRQ & 1.522 & 48.43 & 0.02 & 45.41 & 0.56 & 46.29 & -3.02 \\
J0930.4+8611 & S5 0916+864 & BLL & 0.217 & 45.2 & 0.06 & 43.29 & 0.57 & 43.8 & -1.91 \\
J2056.2-4715 & PKS 2052-47 & FSRQ & 1.489 & 48.33 & 0.02 & 45.34 & 0.56 & 46.26 & -2.99 \\
J2139.3-4236 & MH 2136-428 & BLL & 0.092 & 45.11 & 0.02 & 43.23 & 0.56 & 42.74 & -1.88 \\
J0641.2+7315 & S5 0633+73 & FSRQ & 1.85 & 47.53 & 0.08 & 44.82 & 0.57 & 45.86 & -2.71 \\
J1727.1+4531 & S4 1726+45 & FSRQ & 0.714 & 46.89 & 0.03 & 44.4 & 0.56 & 45.09 & -2.49 \\
J1728.2+5015 & IZW 187 & BLL & 0.055 & 43.83 & 0.08 & 42.39 & 0.57 & 43.28 & -1.44 \\
J1700.2+6831 & BZQ J1700+6830 & FSRQ & 0.301 & 46.2 & 0.02 & 43.95 & 0.56 & 44.37 & -2.25 \\
J0721.9+7120 & S5 0716+714 & BLL & 0.3 & 46.78 & 0.01 & 44.46 & 0.26 & 44.76 & -2.32* \\
J0710.5+5908 & EXO 0706.1+5913 & BLL & 0.125 & 44.7 & 0.09 & 42.96 & 0.57 & 43.8 & -1.74 \\
J0654.5+5043 & CRATES J0654+5042 & FSRQ & 0.135 & 45.17 & 0.03 & 43.27 & 0.56 & 43.62 & -1.9 \\
J0841.6+7052 & 4C +71.07 & FSRQ & 2.172 & 48.36 & 0.05 & 45.16 & 0.26 & 45.97 & -3.2* \\
J0654.2+4514 & S4 0650+45 & FSRQ & 0.933 & 47.34 & 0.03 & 44.7 & 0.56 & 45.1 & -2.65 \\
J0747.7+4501 & B3 0745+453 & FSRQ & 0.192 & 44.78 & 0.1 & 43.02 & 0.57 & 43.87 & -1.77 \\
J0830.5+2407 & B2 0827+24 & FSRQ & 0.941 & 47.17 & 0.04 & 44.07 & 0.26 & 45.22 & -3.1* \\
J1017.0+3531 & B2 1015+35B & FSRQ & 1.228 & 47.04 & 0.11 & 44.5 & 0.57 & 45.23 & -2.54 \\
J1033.2+4117 & S4 1030+415 & FSRQ & 1.117 & 47.18 & 0.04 & 44.59 & 0.57 & 45.35 & -2.59 \\
J1153.2+4935 & SBS 1150+497 & FSRQ & 0.334 & 45.69 & 0.05 & 43.61 & 0.57 & 45.19 & -2.08 \\
J0710.8+4733 & S4 0707+476 & BLL & 1.292 & 47.25 & 0.06 & 44.64 & 0.57 & 45.59 & -2.62 \\
J1637.7+4714 & 4C +47.44 & FSRQ & 0.74 & 46.88 & 0.03 & 44.39 & 0.57 & 45.42 & -2.49 \\
J1559.0+5627 & BZB J1558+5625 & BLL & 0.3 & 45.62 & 0.05 & 43.56 & 0.57 & 44.3 & -2.05 \\
J1542.9+6129 & BZB J1542+6129 & BLL & 0.124 & 45.45 & 0.02 & 43.45 & 0.56 & 43.19 & -2.0 \\
J1110.2+7134 & BZB J1110+7133 & BLL & 0.201 & 44.61 & 0.12 & 42.9 & 0.58 & 43.82 & -1.71 \\
J0816.5+5739 & SBS 0812+578 & BLL & 0.226 & 45.21 & 0.07 & 43.3 & 0.57 & 44.01 & -1.91 \\
J1136.7+7009 & MKN 180 & BLL & 0.046 & 43.85 & 0.06 & 42.4 & 0.56 & 43.48 & -1.45 \\
J1031.0+5053 & 1ES 1028+511 & BLL & 0.36 & 45.76 & 0.07 & 43.66 & 0.57 & 43.57 & -2.1 \\
J0958.6+6533 & S4 0954+658 & BLL & 0.368 & 45.98 & 0.03 & 43.8 & 0.56 & 44.42 & -2.18 \\
J0921.9+6216 & S4 0917+62 & FSRQ & 1.446 & 47.41 & 0.06 & 44.74 & 0.57 & 45.66 & -2.67 \\
J0809.8+5218 & 1ES 0806+524 & BLL & 0.138 & 45.14 & 0.04 & 43.25 & 0.56 & 43.29 & -1.89 \\
J1033.9+6050 & S4 1030+61 & FSRQ & 1.401 & 47.87 & 0.02 & 45.04 & 0.56 & 45.51 & -2.83 \\
J0945.9+5751 & BZB J0945+5757 & BLL & 0.229 & 45.0 & 0.07 & 43.16 & 0.57 & 43.82 & -1.84 \\
J1420.2+5422 & 1418+546 & BLL & 0.151 & 44.8 & 0.05 & 43.02 & 0.57 & 43.79 & -1.77 \\
J1154.4+6019 & CRATES J1154+6022 & FSRQ & 0.543 & 46.19 & 0.06 & 43.94 & 0.57 & 44.51 & -2.25 \\
J1019.0+5915 & BZB J1018+5911 & BLL & 0.191 & 44.6 & 0.1 & 42.9 & 0.57 & 43.74 & -1.71 \\
J0957.7+5522 & 4C +55.17 & FSRQ & 0.895 & 47.7 & 0.02 & 44.93 & 0.56 & 45.56 & -2.77 \\
J1248.2+5820 & PG 1246+586 & BLL & 0.847 & 47.2 & 0.03 & 44.6 & 0.57 & 44.5 & -2.6 \\
J1151.5+5857 & BZB J1151+5859 & BLL & 0.127 & 44.59 & 0.08 & 42.89 & 0.57 & 43.9 & -1.7 \\
J1058.6+5628 & BZB J1058+5628 & BLL & 0.144 & 45.47 & 0.03 & 43.47 & 0.56 & 43.68 & -2.0 \\
J1037.6+5712 & BZB J1037+5711 & BLL & 0.106 & 44.97 & 0.04 & 43.14 & 0.56 & 43.34 & -1.83 \\
J0929.5+5009 & BZB J0929+5013 & BLL & 0.37 & 45.57 & 0.08 & 43.53 & 0.57 & 44.1 & -2.04 \\
J0818.2+4223 & S4 0814+425 & BLL & 0.245 & 46.17 & 0.02 & 45.01 & 0.26 & 44.41 & -1.17* \\
J1253.1+5302 & S4 1250+53 & BLL & 0.204 & 45.69 & 0.03 & 43.61 & 0.56 & 44.17 & -2.08 \\
J0903.4+4651 & S4 0859+470 & FSRQ & 1.462 & 46.95 & 0.11 & 44.44 & 0.58 & 46.1 & -2.51 \\
J0834.3+4221 & S4 0830+42 & FSRQ & 0.249 & 45.24 & 0.07 & 43.32 & 0.57 & 44.18 & -1.93 \\
J0920.9+4441 & S4 0917+44 & FSRQ & 2.189 & 48.72 & 0.02 & 45.6 & 0.56 & 45.84 & -3.12 \\
J1439.2+3932 & PG 1437+398 & BLL & 0.349 & 45.31 & 0.13 & 43.36 & 0.58 & 44.26 & -1.95 \\
J1012.5+4227 & BZB J1012+4229 & BLL & 0.364 & 45.27 & 0.13 & 43.34 & 0.58 & 44.15 & -1.94 \\
J0948.8+4040 & 4C +40.24 & FSRQ & 1.252 & 46.98 & 0.09 & 43.71 & 0.28 & 45.65 & -3.28* \\
J0917.0+3900 & S4 0913+391 & FSRQ & 1.269 & 46.95 & 0.1 & 44.44 & 0.57 & 45.73 & -2.51 \\
J0337.0+3200c & NRAO 140 & FSRQ & 1.258 & 47.44 & 0.07 & 44.76 & 0.57 & 45.54 & -2.68 \\
J0339.4-0144 & PKS 0336-019 & FSRQ & 0.852 & 46.96 & 0.04 & 43.93 & 0.26 & 45.31 & -3.03* \\
J0340.6-2113 & PKS 0338-214 & BLL & 0.048 & 43.61 & 0.07 & 42.24 & 0.57 & 43.56 & -1.37 \\
J0405.8-1309 & PKS 0403-13 & FSRQ & 0.571 & 45.95 & 0.1 & 43.78 & 0.57 & 45.68 & -2.17 \\
J0424.7+0034 & PKS 0422+004 & BLL & 0.31 & 45.88 & 0.04 & 43.73 & 0.57 & 43.88 & -2.14 \\
J0423.2-0120 & PKS 0420-01 & FSRQ & 0.915 & 47.62 & 0.02 & 45.22 & 0.26 & 45.24 & -2.4* \\
J0428.6-3756 & PKS 0426-380 & BLL & 1.11 & 48.37 & 0.01 & 45.37 & 0.56 & 45.53 & -3.0 \\
J0448.9+1121 & 0446+112 & FSRQ & 1.207 & 47.73 & 0.03 & 44.95 & 0.57 & 45.22 & -2.78 \\
J0442.7-0017 & PKS 0440-00 & FSRQ & 0.844 & 47.57 & 0.01 & 44.84 & 0.56 & 45.41 & -2.72 \\
J0501.2-0155 & 4C -02.19 & FSRQ & 2.286 & 48.02 & 0.07 & 45.28 & 0.27 & 46.3 & -2.74* \\
J0505.5+0501 & PKS 0502+049 & FSRQ & 0.954 & 46.99 & 0.07 & 44.46 & 0.57 & 45.16 & -2.53 \\
J0530.8+1333 & PKS 0528+134 & FSRQ & 2.07 & 48.32 & 0.05 & 45.36 & 0.27 & 45.92 & -2.96* \\
J0532.7+0733 & TEX 0529+075 & FSRQ & 1.254 & 47.78 & 0.03 & 44.98 & 0.57 & 45.57 & -2.8 \\
J0538.8-4405 & PKS 0537-441 & BLL & 0.896 & 48.25 & 0.01 & 45.29 & 0.56 & 45.53 & -2.96 \\
J0608.0-0836 & OH -10 & FSRQ & 0.872 & 47.03 & 0.05 & 44.02 & 0.27 & 45.42 & -3.01* \\
J0738.0+1742 & PKS 0735+17 & BLL & 0.424 & 46.57 & 0.02 & 44.19 & 0.56 & 44.47 & -2.38 \\
J0739.2+0138 & PKS 0736+01 & FSRQ & 0.191 & 45.65 & 0.02 & 42.91 & 0.26 & 44.1 & -2.74* \\
J0730.2-1141 & PKS 0727-11 & FSRQ & 1.591 & 48.75 & 0.01 & 45.62 & 0.56 & 44.8 & -3.13 \\
J0849.0+0455 & BZB J0849+0455 & BLL & 1.07 & 46.6 & 0.14 & 44.21 & 0.58 & 45.37 & -2.39 \\
J2319.1-4208 & PKS 2316-423 & BLL & 0.056 & 43.54 & 0.19 & 42.2 & 0.59 & 44.02 & -1.34 \\
J2204.6+0442 & PKS 2201+04 & BLL & 0.028 & 43.11 & 0.08 & 41.91 & 0.56 & 43.38 & -1.2 \\
J2159.9+1023 & CRATES J2200+1030 & BLL & 0.172 & 44.62 & 0.12 & 42.91 & 0.57 & 44.23 & -1.71 \\
J0250.6+1713 & BZQ J0250+1712 & FSRQ & 0.207 & 45.13 & 0.08 & 43.25 & 0.57 & 44.25 & -1.89 \\
J0112.8+3208 & NRAO 62 & FSRQ & 0.603 & 46.99 & 0.02 & 44.46 & 0.56 & 45.21 & -2.52 \\
J0909.7-0229 & PKS 0907-023 & FSRQ & 0.957 & 47.13 & 0.04 & 44.55 & 0.57 & 45.3 & -2.57 \\
J0912.9-2102 & BZB J0913-2103 & BLL & 0.198 & 45.01 & 0.09 & 43.16 & 0.57 & 44.45 & -1.84 \\
J1204.3-0711 & BZB J1204-0710 & BLL & 0.185 & 44.91 & 0.08 & 43.1 & 0.57 & 43.84 & -1.81 \\
J1317.9+3426 & B2 1315+34A & FSRQ & 1.05 & 46.53 & 0.12 & 44.16 & 0.58 & 45.55 & -2.37 \\
J1312.4-2157 & PKS 1309-216 & BLL & 1.489 & 47.55 & 0.05 & 44.83 & 0.57 & 45.84 & -2.72 \\
J1351.1+0032 & PKS 1348+007 & FSRQ & 2.084 & 47.64 & 0.08 & 44.89 & 0.57 & 45.96 & -2.75 \\
J1435.1+2022 & CRATES J1435+2021 & BLL & 0.227 & 45.01 & 0.09 & 43.16 & 0.57 & 44.61 & -1.84 \\
J1744.1+1934 & 1ES 1741+196 & BLL & 0.083 & 44.18 & 0.12 & 42.62 & 0.57 & 43.81 & -1.56 \\
J1849.4+6706 & 4C +66.20 & FSRQ & 0.657 & 47.25 & 0.02 & 44.63 & 0.56 & 45.39 & -2.61 \\
J0611.8-6059 & BZQ J0610-6058 & FSRQ & 0.617 & 46.1 & 0.09 & 43.88 & 0.57 & 45.17 & -2.22 \\
J1219.2+7107 & S5 1217+71 & FSRQ & 0.451 & 45.55 & 0.09 & 43.52 & 0.57 & 44.66 & -2.03 \\
J1136.3+6736 & BZB J1136+6737 & BLL & 0.135 & 44.58 & 0.1 & 42.88 & 0.57 & 43.54 & -1.7 \\
J1454.4+5123 & BZB J1454+5124 & BLL & 1.083 & 47.17 & 0.05 & 44.58 & 0.57 & 45.27 & -2.59 \\
J1203.2+6030 & SBS 1200+608 & BLL & 0.066 & 43.95 & 0.07 & 42.47 & 0.57 & 43.45 & -1.48 \\
J1053.6+4928 & BZB J1053+4929 & BLL & 0.14 & 44.77 & 0.08 & 43.01 & 0.57 & 43.68 & -1.76 \\
J1433.8+4205 & B3 1432+422 & FSRQ & 1.24 & 46.85 & 0.09 & 44.37 & 0.57 & 45.34 & -2.48 \\
J0315.8-1024 & PKS 0313-107 & FSRQ & 0.335 & 45.52 & 0.08 & 43.5 & 0.57 & 44.88 & -2.02 \\
J0612.8+4122 & S4 0609+41 & BLL & 0.097 & 44.98 & 0.03 & 43.14 & 0.56 & 43.74 & -1.83 \\
J0630.9-2406 & BZB J0630-2406 & BLL & 0.062 & 44.57 & 0.04 & 42.88 & 0.56 & 43.2 & -1.69 \\
\hline
\\
\multicolumn{10}{l}{\textbf{Notes:} } \\ 
\multicolumn{10}{l}{(a) Uncertainty on $L^{\rm iso}$ in dex. } \\ 
\multicolumn{10}{l}{(b) Uncertainty on the collimation-corrected $\gamma$-ray luminosity $L$ in dex.} \\ 
\multicolumn{10}{l}{(c) The uncertainty on $P_{\rm jet}$ is 0.7 dex as described in the text. } \\ 
\multicolumn{10}{l}{(d) Values with a ``*'' correspond to those measured by \citen{spush09}, otherwise they were estimated from the best-fit in Fig. \ref{fig:beaming}.} \\ 
\label{tab:agn}
\end{longtable}
\end{landscape}

\clearpage

\begin{landscape}
\begin{longtable}{ccccccccccc}
\multicolumn{11}{l}{{\bf Table~S2.} Properties of the GRB sample.} \\
\\
\hline\hline
Name & $z$ & $t_{90}$ & $\theta_j$ $^{\rm a}$ & $\log E_\gamma^{\rm iso}$ $^{\rm b}$ & $\log L^{\rm iso}$ $^{\rm b}$ & $\log L$$^{\rm c}$ & $\log E_k^{\rm iso}$ $^{\rm d}$ & $\log P_{\rm jet}$$^{\rm e}$ & Observatory & Ref. \\
 &  & (s) & ($^\circ$) & (erg) & (erg s$^{-1}$) & (erg s$^{-1}$) & (erg) & (erg s$^{-1}$) &  &  \\
 \hline
090323 & 3.57 & 133.1 & 2.8 & 54.53 & 53.06 & 50.14 & 54.07 & 49.68 & LAT & 1 \\
090328 & 0.74 & 57.0 & 4.2 & 52.99 & 51.47 & 48.9 & 53.91 & 49.83 & LAT & 1 \\
090902B & 1.82 & 21.0 & 3.9 & 54.5 & 53.63 & 51.0 & 53.75 & 50.24 & LAT & 1 \\
090926A & 2.11 & 20.0 & 9.0 & 54.27 & 53.46 & 51.55 & 52.91 & 50.19 & LAT & 1 \\
081222 & 2.77 & 5.8 & 2.8 & 53.47 & 53.28 & 50.35 & 54.12 & 51.0 & GBM & 2 \\
090424 & 0.54 & 20.2 & 9.8 & 52.64 & 51.52 & 49.69 & 54.22 & 51.26 & GBM & 2 \\
090618 & 0.54 & 24.0 & 6.7 & 53.38 & 52.19 & 50.02 & 53.49 & 50.13 & GBM & 2 \\
091020 & 1.71 & 65.0 & 6.9 & 53.03 & 51.65 & 49.51 & 53.71 & 50.19 & GBM & 2 \\
091127 & 0.48 & 68.7 & 5.5 & 52.14 & 50.47 & 48.13 & 53.36 & 49.35 & GBM & 2 \\
091208B & 1.06 & 71.0 & 7.3 & 52.33 & 50.8 & 48.7 & 53.7 & 50.07 & GBM & 2 \\
970228 & 0.7 & 56.0 & 4.0 & 52.15 & 50.63 & 48.02 & 53.24 & 49.11 & \textit{BeppoSAX} & 3,4,5,22 \\
970508 & 0.84 & 14.0 & 21.6 & 51.74 & 50.86 & 49.71 & 52.96 & 50.92 & BATSE & 3,4,5,22 \\
970828 & 0.96 & 160.0 & 7.1 & 53.34 & 51.43 & 49.32 & 53.57 & 49.55 & BATSE & 3,4,5,22 \\
971214 & 3.42 & 30.0 & 5.5 & 53.32 & 52.49 & 50.16 & 53.89 & 50.73 & BATSE & 3,4,5,22 \\
980613 & 1.1 & 42.0 & 12.6 & 51.73 & 50.43 & 48.81 & 53.05 & 50.13 & \textit{BeppoSAX} & 3,4,5,22 \\
980425 & 0.01 & 23.3 & 11.0 & 48.2 & 46.84 & 45.1 & 49.08 & 45.98 & \textit{BeppoSAX} & 10,8,6 \\
980703 & 0.97 & 76.0 & 11.2 & 52.78 & 51.19 & 49.47 & 53.35 & 50.04 & BATSE & 3,4,5,22 \\
990123 & 1.6 & 61.0 & 4.9 & 54.16 & 52.79 & 50.36 & 54.27 & 50.47 & BATSE & 3,4,5,22 \\
990510 & 1.62 & 57.0 & 3.4 & 53.25 & 51.91 & 49.14 & 54.08 & 49.98 & BATSE & 3,4,5,22 \\
990705 & 0.84 & 32.0 & 5.3 & 53.41 & 52.17 & 49.8 & 52.49 & 48.89 & BATSE & 3,4,5,22 \\
991216 & 1.02 & 15.0 & 4.6 & 53.73 & 52.86 & 50.36 & 54.53 & 51.16 & BATSE & 3,4,5,22 \\
021004 & 0.85 & 9.0 & 6.8 & 53.23 & 52.54 & 50.39 & 52.66 & 49.83 & \textit{BeppoSAX} & 3,4,5,22 \\
000926 & 2.04 & 1.3 & 6.2 & 53.45 & 53.82 & 51.58 & 53.96 & 52.09 & \textit{BeppoSAX} & 3,4,5,22 \\
010222 & 1.48 & 74.0 & 3.2 & 53.93 & 52.46 & 49.65 & 54.32 & 50.04 & \textit{BeppoSAX} & 3,4,5,22 \\
011211 & 2.14 & 51.0 & 6.4 & 52.83 & 51.62 & 49.41 & 53.08 & 49.66 & \textit{BeppoSAX} & 3,4,5,22 \\
020405 & 0.7 & 40.0 & 7.8 & 52.86 & 51.49 & 49.45 & 53.63 & 50.22 & \textit{BeppoSAX} & 3,4,5,22 \\
020813 & 1.25 & 89.0 & 3.1 & 53.89 & 52.29 & 49.47 & 54.31 & 49.89 & HETE & 3,4,5,22 \\
021004 & 2.32 & 52.4 & 12.7 & 52.75 & 51.55 & 49.93 & 53.89 & 51.07 & HETE & 3,4,5,22 \\
031203 & 0.1 & 40.0 & 9.0 & 50.99 & 49.43 & 47.52 & 51.14 & 47.67 & Integral & 9,6 \\
030329 & 0.17 & 22.3 & 5.1 & 52.1 & 50.82 & 48.42 & 52.8 & 49.12 & HETE & 7,11 \\
050709 & 0.16 & 0.1 & 17.2 & 50.0 & 51.22 & 49.87 & 51.05 & 50.92 & BAT & 12 \\
050820A & 2.62 & 26.0 & 6.6 & 53.98 & 53.13 & 50.95 & 54.73 & 51.69 & BAT & 13,14 \\
050904 & 6.3 & 225.0 & 8.0 & 54.12 & 52.63 & 50.62 & 53.95 & 50.45 & BAT & 13,15 \\
060218 & 0.03 & 2100.0 & 80.4 & 49.4 & 46.09 & 46.01 & 49.11 & 45.73 & BAT & 16,23 \\
060418 & 1.49 & 52.0 & 22.5 & 52.6 & 51.28 & 50.16 & 52.09 & 49.65 & BAT & 13 \\
070125 & 1.55 & 60.0 & 13.2 & 53.98 & 52.61 & 51.03 & 52.81 & 49.86 & BAT & 13,17,18 \\
080319B & 0.94 & 50.0 & 7.0 & 52.67 & 51.26 & 49.13 & 54.13 & 50.59 & BAT & 21,13 \\
050505 & 4.27 & 136.0 & 2.0 & 53.27 & 51.86 & 48.66 & 54.77 & 50.16 & BAT & 2 \\
050814 & 5.3 & 48.0 & 2.4 & 53.24 & 52.36 & 49.32 & 54.92 & 51.0 & BAT & 2 \\
051109A & 2.35 & 360.0 & 3.4 & 52.82 & 50.79 & 48.02 & 54.23 & 49.43 & BAT & 2 \\
051221A & 0.55 & 8.0 & 11.8 & 51.41 & 50.7 & 49.02 & 52.64 & 50.25 & BAT & 2 \\
060124 & 2.3 & 3.3 & 2.4 & 53.6 & 53.6 & 50.55 & 54.94 & 51.89 & BAT & 2 \\
060614 & 0.12 & 6.9 & 11.6 & 50.52 & 49.73 & 48.05 & 52.23 & 49.75 & BAT & 2 \\
060707 & 3.42 & 210.0 & 7.9 & 52.85 & 51.18 & 49.16 & 54.01 & 50.31 & BAT & 2 \\
060814 & 0.84 & 1.2 & 3.5 & 52.78 & 52.96 & 50.24 & 53.44 & 50.9 & BAT & 2 \\
061021 & 0.35 & 79.0 & 8.6 & 51.52 & 49.75 & 47.8 & 52.79 & 49.07 & BAT & 2 \\
061222A & 2.09 & 16.0 & 2.7 & 53.48 & 52.76 & 49.8 & 55.36 & 51.68 & BAT & 2 \\
070306 & 1.5 & 3.0 & 4.4 & 52.78 & 52.7 & 50.16 & 53.83 & 51.21 & BAT & 2 \\
070318 & 0.84 & 0.4 & 8.8 & 52.03 & 52.69 & 50.76 & 53.92 & 52.65 & BAT & 2 \\
070508 & 0.82 & 23.4 & 3.5 & 52.87 & 51.76 & 49.03 & 53.03 & 49.19 & BAT & 2 \\
080310 & 2.43 & 32.0 & 3.6 & 52.78 & 51.81 & 49.1 & 53.47 & 49.79 & BAT & 2 \\
080413B & 1.1 & 1.0 & 6.0 & 52.29 & 52.61 & 50.35 & 54.14 & 52.2 & BAT & 2 \\
090313 & 3.38 & 170.0 & 3.1 & 52.82 & 51.23 & 48.38 & 54.38 & 49.94 & BAT & 2 \\
091018 & 0.97 & 106.5 & 4.7 & 51.77 & 50.04 & 47.57 & 53.08 & 48.87 & BAT & 2 \\
\hline
090423$^{\rm f}$ & 8.2 & 10.3 & $>12$ & 53 & 53 & $>51.3$ & 53.6 & $>51.9$ & BAT & 19,20 \\
020903$^{\rm g}$ & 0.251 & 10 & $<90$ & 49.04 & 48.14 & $<48.14$ & 50.6 & $<49.7$ & HETE-2 & 24,25 \\
\hline
\\
\multicolumn{11}{l}{\textbf{Notes:} } \\ 
\multicolumn{11}{l}{(a) Uncertainty of 0.1 dex as described in the text. } \\ 
\multicolumn{11}{l}{(b) Uncertainty of 0.2 dex as described in the text.} \\ 
\multicolumn{11}{l}{(c) Uncertainty of 0.3 dex as derived from uncertainties in $\theta_j$ and $E_\gamma^{\rm iso}$.} \\ 
\multicolumn{11}{l}{(d) Uncertainty of 0.5 dex as described in the text.} \\ 
\multicolumn{11}{l}{(e) Uncertainty of 0.54 dex as derived from uncertainties in $\theta_j$ and $E_k^{\rm iso}$.} \\ 
\multicolumn{11}{l}{(f) Lower limit on the opening angle. Not taken into account in the statistics. } \\ 
\multicolumn{11}{l}{(g) X-ray flash with an upper limit on the opening angle. Not taken into account in the statistics.} \\ 
\multicolumn{11}{l}{\textbf{References:} 1. Cenko et al. 2012 \cite{scenko12}, 2. Racusin et al. 2011 \cite{sracusin11}, 3. Bloom et al. 2003 \cite{sbloom03}, } \\ 
\multicolumn{11}{l}{4. Berger et al. 2003 \cite{sberger03}, 5. Lloyd-Ronning \& Zhang 2004 \cite{slz04}, 6. Ghisellini et al. 2006 \cite{sguise06},  } \\
\multicolumn{11}{l}{7. Berger et al. 2003b \cite{sberger03nat}, 8. Li \& Chevalier 1999 \cite{sli99}, 9. Soderberg et al. 2004 \cite{ssode04}, } \\ 
\multicolumn{11}{l}{10. Yamasaki et al. 2003 \cite{syama03}, 11. Vanderspek et al. 2004 \cite{svander04}, 12. Villasenor et al. 2005 \cite{svilla05}, } \\ 
\multicolumn{11}{l}{13. Cenko et al. 2010 \cite{scenko10}, 14. Cenko et al. 2006 \cite{scenko06}, 15. Kawai et al. 2006 \cite{skawai06}, } \\ 
\multicolumn{11}{l}{16. Campana et al. 2006 \cite{scampana06}, 17. Racusin et al. 2007 \cite{sracusin07}, 18. Bellm et al. 2008 \cite{sbellm08}, } \\
\multicolumn{11}{l}{19. Tanvir et al. 2009 \cite{stanvir09}, 20. Chandra 2010 \cite{schandra10}, 21. Racusin et al. 2008 \cite{sracusin08}, } \\
\multicolumn{11}{l}{22. Fan \& Piran 2006 \cite{sfan06}, 23. Soderberg et al. 2006 \cite{ssode06}, 24. Soderberg et al. 2004b \cite{ssode04apj}, } \\
\multicolumn{11}{l}{25. Sakamoto et al. 2004 \cite{ssaka04}.}
\label{tab:grb}
\end{longtable}
\end{landscape}

\end{document}